\documentclass[12pt,preprint]{aastex}
\def\be{\begin{equation}}
\def\ee{\end{equation}}
\def\bea{\begin{eqnarray}}
\def\eea{\end{eqnarray}}
\begin{document}
\title{GRB Radiative Efficiencies Derived from the {\em
Swift} Data: \\ GRBs vs. XRFs, Long vs. Short}

\author{Bing Zhang$^1$, Enwei Liang$^{1,2}$, Kim L. Page$^{3}$,  Dirk
Grupe$^{4}$, Bin-Bin Zhang$^{5,1}$, Scott D. Barthelmy$^{6}$,  David
N. Burrows$^{4}$, Sergio
Campana$^6$, Guido Chincarini$^{7,8}$, Neil Gehrels$^{6}$, Shiho
Kobayashi$^{9}$, Peter M\'{e}sz\'{a}ros$^{4,10}$, Alberto Moretti$^7$,
John A. Nousek$^{4}$, Paul T. O'Brien$^3$, Julian P. Osborne$^3$,
Peter W. A. Roming$^4$, Takanori Sakamoto$^{6}$, Patricia Schady$^4$,
Richard Willingale$^{3}$}
\affil{$^1$ Department of Physics, University of Nevada, Las Vegas, NV
89154; \\bzhang@physics.unlv.edu, lew@physics.unlv.edu\\
$^2$Department of Physics, Guangxi University, Nanning 530004, China\\
$^3$ Department of Physics and Astronomy, University of Leicester,
Leicester LE1 7RH, UK. \\
$^4$ Department of Astronomy and Astrophysics, Pennsylvania State
University, 525 Davey Laboratory, University Park, PA 16802.\\
$^5$ National Astronomical Observatory/Yunnan Observatory, Chinese
Academy of Sciences, Kunming 650011, China \\
$^6$ NASA/Goddard Space Flight Center, Greenbelt, MD 20771. \\
$^7$ INAF-Osservatorio Astronomico di Brera, Via Bieanchi 46, I-23807
Merate, Italy. \\
$^8$ Univerit\'a degli studi Milano-Bicocca, Dipartmento di Fisica,
Piazza delle Scienze 3, I-20126 Milan, Italy. \\
$^9$ Astrophysics Research Institute, Liverpool John Moores
University, Twelve Quays House, Birkenhead, CH41 1LD, UK. \\
$^10$ Department of Physics, Pennsylvania State
University, 104 Davey Laboratory, University Park, PA 16802.\\
}

\begin{abstract}
We systematically analyze the prompt emission and the early afterglow
data of a sample of 31 GRBs detected by {\em Swift} before September
2005, and estimate the GRB radiative efficiency.  BAT's
narrow band inhibits a precise determination of the GRB spectral
parameters, and we have developed a method to estimate these
parameters with the hardness ratio information. The shallow decay
component commonly existing in early X-ray afterglows, if interpreted
as continuous energy injection in the external shock, suggests that
the GRB efficiency previously derived from the late-time X-ray data
were not reliable. We calculate two radiative efficiencies using the
afterglow kinetic energy $E_K$ derived at the putative deceleration
time ($t_{\rm dec}$) and at the break time ($t_b$) when the energy
injection phase ends, respectively. At $t_b$ XRFs appear to
be less efficient than normal GRBs. However, when we analyze the data 
at $t_{dec}$ XRFs are found to be as efficient as GRBs.
Short GRBs have similar radiative efficiencies to long GRBs despite of
their different progenitors. Twenty-two bursts in the sample are
identified to have the afterglow cooling frequency below the X-ray
band. Assuming $\epsilon_e = 0.1$, we find $\eta_\gamma
(t_b)$ usually $<10\%$ and $\eta_\gamma (t_{dec})$
varying from a few percents to $> 90\%$. Nine GRBs in the sample have
the afterglow cooling frequency above the X-ray band for a very long
time. This suggests a very small $\epsilon_B$ and/or a very low
ambient density $n$.
\end{abstract}


%%%%%%%%%%%%%%%%%%%%%%%%%%%%%%%%%%%%%%%%%%%%%%%%%%%%%%%%%%%%%%%%%%%%%

\section{Introduction}
Gamma-ray bursts (GRBs) are believed to be the most luminous
electromagnetic explosions in the universe.  These erratic, transient
events in $\gamma$-rays are followed by long-lived, decaying
afterglows in longer wavelengths.  The widely accepted model of this
phenomenon is the fireball model (M\'esz\'aros 2002;
Zhang \& M\'esz\'aros 2004; Piran 2005), which depicts the observed
prompt gamma-ray emission as the synchrotron emission from the
internal shocks in an erratic, unsteady, relativistic fireball (Rees \&
M\'esz\'aros 1994), and interprets the broadband afterglow emission as
the synchrotron emission from an external shock that expands into the
circumburst medium (M\'{e}sz\'{a}ros \& Rees 1997; Sari et
al. 1998). The GRB radiative efficiency, which is defined as
\be
\eta_\gamma \equiv \frac{E_\gamma}{E_\gamma+E_{K}},
\label{eta}
\ee
is an essential quantity to understand the nature of the bursts.
Here $E_{\gamma}$ is the isotropic gamma-ray energy\footnote{Notice
that the notation $E_\gamma$ is different from that used in some other
papers (e.g. Frail et al. 2001) that denotes the geometry-corrected
gamma-ray energy.} and $E_{K}$ is the isotropic kinetic energy of the
fireball right after the prompt 
gamma-ray emission is over. It gives a direct measure of how
efficient the burster dissipates the total energy into radiation
during the GRB prompt emission phase.

In the pre-{\em Swift} era, only the late time fireball kinetic energy
$E_K$ was derived or estimated using the late time afterglow data. Two
methods have been proposed. The most adequate one is through broadband
afterglow modeling (e.g. Panaitescu \& Kumar 2001).  This method
requires well-sampled multi-wavelength afterglow data. The method thus
can only be applied to a small sample of GRBs. A more convenient
method is to use the X-ray afterglow data alone (Freedman \& Waxman
2001; Berger et al. 2003; Lloyd-Ronning \& Zhang 2004). At a late
enough epoch (e.g. 10 hours after the burst trigger), 
the X-ray band is likely above the cooling frequency, so that the
X-ray flux gives a good measure of the degenerate quantity $\epsilon_e
E_K$, where $\epsilon_e$ is the fraction of the electron energy in the
internal energy of the shock. If $\epsilon_e$ could be estimated, one
can then derive $E_K$ directly from the X-ray data.  Assuming a
simple extrapolation of the late time light curve to earlier epochs,
Lloyd-Ronning \& Zhang (2004) took into account the fireball radiative
loss correction to estimate $E_{K}$ right after the prompt emission
phase, and estimated the efficiency of 17 GRBs/XRFs observed in the
pre-{\em Swift} era. They discovered a shallow positive correlation
between $\eta_\gamma$ and $E_\gamma$ or $E_p$. According to this
shallow correlation, softer, under-luminous bursts (e.g. X-ray
flashes) tend to have a lower radiative efficiency. Similar
conclusions were also drawn by Lamb et al (2005).

These previous GRB efficiency studies employing the late afterglow
data inevitably introduce some uncertainties on the $E_{K}$
measurements, including the possible corrections of radiative fireball
energy loss and additional energy injection in the early afterglow
phase. In order to reduce these uncertainties very early afterglow
observations are 
needed. The successful operation of NASA's {\em Swift} GRB mission
(Gehrels et al. 2004) makes this possible.  Very early X-ray afterglow
data for a large number of GRBs have been recorded by the X-Ray
Telescope (XRT) on board {\em Swift} (Burrows et al. 2005a). A large
sample of early X-ray afterglow data have
been collected, typically 100 seconds after the triggers. These
observations indeed show novel, unexpected behaviors in the early
afterglow phase (Tagliaferri et al. 2005; Burrows et al. 2005b; Nousek
et al. 2006; Zhang et al. 2006; O'Brien et al. 2006), and make it
feasible to more robustly estimate $E_{K}$ and hence, $\eta_\gamma$.

The XRT light curves of many bursts could be synthesized to a
canonical light curve that is composed of 5 components (Zhang et
al. 2006), an early rapid decay component consistent with the tail of
the prompt emission, a frequently-seen shallow-than-normal decay
component likely originated from a refreshed external forward shock, a
normal decay component due to a freely expanding fireball, an
occasionally-seen post jet break segment, as well as erratic X-ray
flares harboring in nearly half of {\em Swift} GRBs that are likely
due to reactivation of the GRB central engine. The most relevant
segment for the efficiency problem is the shallow decay component
(Zhang et al. 2006; Nousek et al.  2006). Most of the {\em Swift}/XRT
afterglow light curves in our sample have such an early shallow decay
segment. The origin of this segment is currently not identified. If it
is due to continuous energy injection, the initial afterglow energy
$E_K$ must be significantly smaller than estimated using the late time
data. The previous efficiency analysis using late X-ray afterglow data
then tend to overestimate $E_K$, and hence, underestimate
$\eta_\gamma$. The early tight UVOT upper limits for many Swift GRBs
are also likely related to this shallow decay component (Roming et
al. 2006).  It is therefore of great interest to revisit the
efficiency problem using the very early XRT data.

X-ray flashes (XRFs, Heise et al. 2001; Kippen et al. 2002) naturally
extend long-duration GRBs into the softer and fainter regime
(e.g. Lamb et al. 2005; Sakamoto et al. 2006). It is now known that
the softness of the bursts is not due to their possible high redshifts
(Soderberg et al. 2004, 2005).  The remaining possibilities include
either extrinsic (e.g. viewed at different viewing angles, Yamazaki et
al. 2002, 2004; Zhang et al. 2004a,b; Liang \& Dai 2004a; Huang et
al. 2004), or intrinsic (e.g. different burst parameters such
as Lorentz factor, luminosity, etc, Dermer et al. 1999; Kobayashi et
al. 2002; M\'esz\'aros et al. 2002; Zhang \& M\'esz\'aros 2002c; Huang
et al. 2002; Rees \& M\'esz\'aros 2005; Barraud et al. 2005) reasons.
The radiative efficiency of XRFs may provide a clue to identify the
correct mechanism.  Previous analyses of late time X-ray data
suggest that the XRFs typically have lower radiative efficiencies
(Soderberg et al. 2004; Lloyd-Ronning \& Zhang 2004; Lamb et
al. 2005). It is desirable to investigate whether this is still true
with the early afterglow data.

Recently, afterglows of several short-duration GRBs have been detected
(Gehrels et al.  2005; Fox et al. 2005; Villasenor et al. 2005; Hjorth
et al. 2005; Barthelmy et al.  2005a; Berger et al. 2005c). The data
suggest that they are distinct from long GRBs and very likely have
different progenitor systems. It is therefore of great interest to
explore the radiative efficiencies of short-hard GRBs (SHGs) and
compare them with those of long GRBs.

In this paper we systematically analyze the prompt emission
and the early afterglow data for a sample of 31 GRBs detected by {\em
Swift} before September 2005 through reducing the Burst Alert
Telescope (BAT, Barthelmy et al. 2005b) and XRT data. We present the
sample and the BAT/XRT
data analysis methods in \S\ref{sec:data}. The gamma-to-X fluence ratio
($R_{\gamma/X}$) is an observation-defined apparent GRB efficiency
indicator. We perform statistical analysis of this parameter in
\S\ref{sec:rel-eff}. In \S\ref{sec:model}, we perform more detailed
theoretical modeling to estimate $E_{K}$ (\S\ref{sec:EK}) and
$\eta_\gamma$ (\S\ref{sec:eta}). Our results are summarized in
\S\ref{sec:conclusion} with some discussion. Throughout the paper
the cosmological parameters $H_0 = 71$ km s$^{-1}$ Mpc$^{-1}$,
$\Omega_M=0.3$, and $\Omega_\Lambda=0.7$ have been adopted.

%%%%%%%%%%%%%%%%%%%%%%%%%%%%%%%%%%%%%%%%%%%%%%%%%%%%%%%%%%%%%%%%%%%%%

\section{Data}
\label{sec:data}

Our sample includes 31 GRBs observed by {\em Swift} before
September 1, 2005. The BAT-XRT joint early X-ray afterglow light
curves of these bursts and the detailed data reduction procedures have
been presented in O'Brien et al. (2006).

\subsection{Prompt Gamma-rays}
\label{sec:BAT}

The BAT data have been processed using the standard BAT analysis
software ({\em Swift} software v. 2.0).  The BAT-band
gamma-ray fluence $S_{\gamma,obs}$ could be directly derived from the
data. For the purpose of estimating GRB efficiency, on the other hand,
one needs to estimate the total energy output of the GRB, which
requires to extrapolate $S_{\gamma,obs}$ to a wider bandpass to get
$S_\gamma$ ($1-10^4$ keV adopted in this paper). This requires the
knowledge of the spectral parameters of the prompt emission.

It is well known that the GRB spectrum is typically fitted by a Band
function (Band et al. 1993), which is a smoothly-joint-broken power
law characterized by two photon indices $\Gamma_1$ and $\Gamma_2$
(with the convention $N(E) d E \propto E^{\Gamma} dE$ adopted
throughout the text) and a break energy $E_0$. The peak energy of the
$\nu f_\nu$ spectrum is $E_p=(2+\Gamma_1)E_0$. In order to derive
these parameters, the observed spectrum of a burst should cover the
energy band around $E_0$. The spectra of both long and short GRBs
observed by CGRO/BATSE (covering 20-2000 keV) are well fitted by the
Band-function, with the typical values of $\Gamma_1\sim -1$,
$\Gamma_2\sim -2.3$, and $E_p\sim 250$ keV (Preece et al. 2000;
Ghirlanda et al. 2003). The spectra of XRFs could be also fitted by
the Band-function, typically with $\Gamma_1\sim -1$ and a lower
$\Gamma_2$ than typical GRBs (Lamb et al. 2005, Sakamoto et al. 2005,
2006; Cui et al.  2005). BAT has a narrower energy band (15-150 keV)
than BATSE and HETE-2. The typical $E_p$ of the bright BATSE sample is
well above the BAT band. BAT's observations therefore cannot well
constrain $E_p$ and $\Gamma_2$ for many GRBs. Most observed spectra by
BAT are well fitted by a simple power law. However, this is due to the
intrinsic limitation of the instrument. Four GRBs in our sample, i.e. GRBs
050401, 050525A, 050713A, and 050717, were simultaneously observed by
{\em Konus-Wind} (with an energy band of 20-2000 keV). The spectra of
these bursts could be also well fitted by a Band-function or a cutoff
power law spectrum (Golenetskii et al. 2005a-d, Krimm et al. 2006). We
assume that the broad band spectra of all the bursts in our sample
could be fitted by the Band-function and then make the corrections to the
observed fluences. Our procedure to derive spectral parameters are as
follows. For more detailed data analysis procedure and burst
parameters, we refer the readers to the {\em Swift} first catalog
paper (L. Angelini et al. 2006, in preparation).

We first fit an observed spectrum with a Band-function, a power law
with exponential cutoff, and a simple power law, respectively. By
comparing reduced $\chi^2$ of these fits, we pick up the best-fit
model among the three. In our sample, only 4 GRBs (050128, 050219A,
050525A and 050716) could be well fitted by a Band function, if one
assigns $\Gamma_2$ in the range of $-5$ to
$-2$\footnote{Although $\Gamma_2\sim -2.3$ for typical GRBs,
$\Gamma_2$ could be as low as $\sim -5$ for XRFs (e.g. Cui et
al. 2005; Sakamoto et al. 2005).}. Due to the great uncertainty of
$\Gamma_2$, the cutoff power law model could also fit these 4 bursts,
with the cutoff energy being in the BAT band. The rest 27 bursts in
the sample are best fitted by a simple power law, with the photon
index $\Gamma^{PL}$ ranging from $\sim -1$ to $\sim -3$ (see Table
1). For the 8 GRBs with the Band-function parameters available (GRBs
050128, 050219A, 050401, 050525A, 050713A, 050716, 050717, and 051221)
either from a fit to the BAT data or from a joint BAT-{\em Konus-Wind}
fit, we make straightforward extrapolation of $S_{\gamma,obs}$ to
derive $S_\gamma$ in the $1-10^4$ keV band. For the rest 24 bursts
whose observed spectra are fitted by a simple power law, the
extrapolation is not straightforward. Generally we employ the hardness
ratio information to place additional constraints to the spectral
parameters. The analyses are carried out on case-to-case base. 
Nonetheless, depending on the value of $\Gamma^{PL}$,
one could crudely group the cases into three categories.

\begin{figure*}
\epsscale{1.0}
\plotone{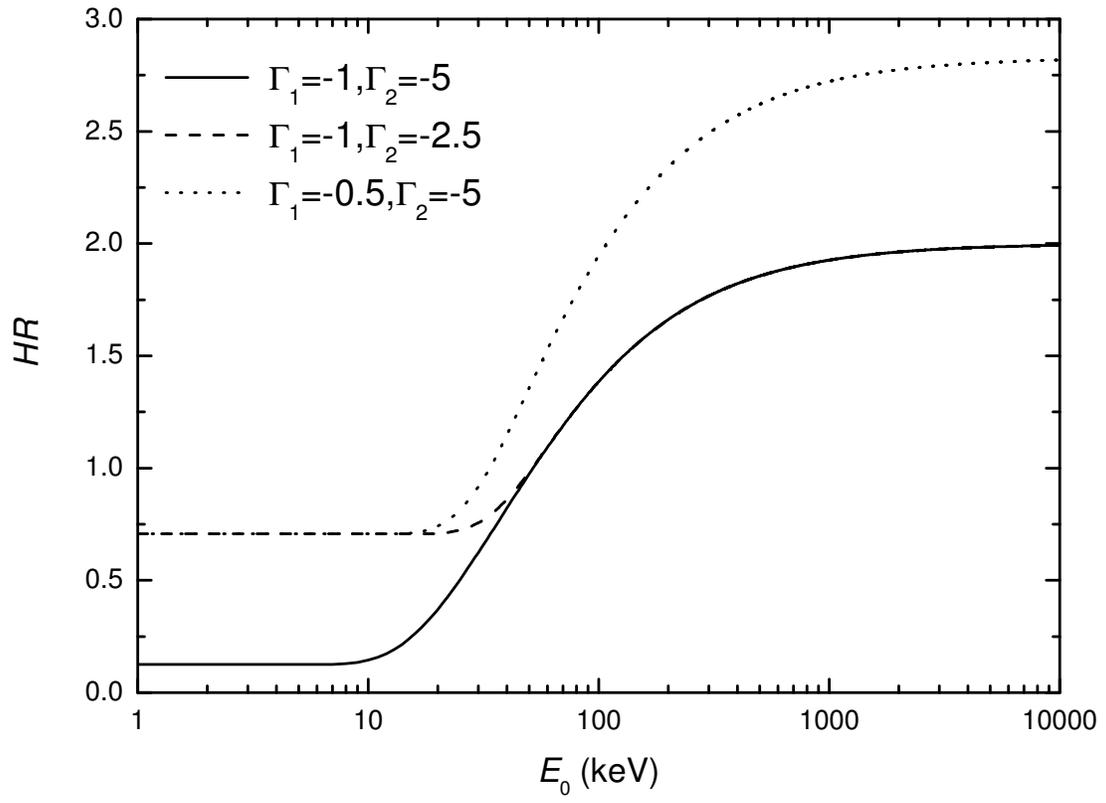}
\caption{The $HR-E_0$ relation for a Band-spectrum. Different sets of
($\Gamma_1$, $\Gamma_2$) have been plotted.}
\label{HR}
\end{figure*}

Case I: $-2.3 \lesssim \Gamma^{PL}\lesssim -1.2$ (16 out of 24 in the
sample). Since for typical bursts in the BATSE sample one has
$\Gamma_1\sim -1$ and $\Gamma_2\sim -2.3$ (Preece et al. 2000), it is
expected that a rough fit to the Band function by a simple power law
would lead to $\Gamma^{PL}$ in this range. The break energy $E_0$ of
these bursts should be within or near the edges of the BAT band. The
hardness ratio ($HR$), which in our analysis is defined as the ratio
of the fluence in the 50-100 keV band to that in the 25-50 keV band,
could be directly measured from the simple PL fit model. 
Theoretically, on the other hand, $HR$ is a function of $\Gamma_1$,
$\Gamma_2$ and $E_0$ for the Band-function model (Cui et al. 2005, see
Fig.\ref{HR}). One can then in principle apply $HR$ as another agent
to constrain the spectral parameters by requiring 
\begin{equation}
HR^{mod}=HR^{obs},
\label{criterion}
\end{equation}
where $HR^{mod}$ and $HR^{obs}$ are the hardness ratio derived from
the Band-function model and from the data, respectively. 
To proceed, we first assign a set of ``standard'' guess
values to the spectral parameters, e.g. $\Gamma_1=-1$,
$\Gamma_2=-2.3$, and $E_p=250$ keV, and then perform a Band-function
fit to the data. We then derive $\Gamma_1$, $\Gamma_2$, and $E_0$ from
the best fit, which usually deviate from the guess values. Generally,
these parameters have very large error bars. We then apply the $HR$
criterion to the results.  If the calculated $HR^{mod}$ using the
best-fit parameters match $HR^{obs}$ well, this set of parameters are
taken. Otherwise, we adjust spectral parameters to achieve the best
match. The process is eased thanks to several properties of the
$HR-E_0$ relation (Fig.\ref{HR}). First, when $E_p$ is high enough
(say, higher than 30 keV), $HR$ is essentially independent on
$\Gamma_2$. Also for $\Gamma_2<-2$, most energy is emitted around
$E_p$ and the extrapolated broad band fluence is insensitive to
$\Gamma_2$. We therefore take the best-fit $\Gamma_2$ value or fix it
to $\sim -2.3$ when $\Gamma_2$ is poorly constrained. $HR^{mod}$
therefore mainly depends on $\Gamma_1$ and $E_0$. Another interesting
feature is that as one decreases $\Gamma_1$ (i.e. softer spectrum),
the corresponding $E_0$ would increase given the same observed
spectrum. The resulting $E_p = (2+\Gamma_1) E_0$, on the other hand,
is not very sensitive to $\Gamma_1$ since the variations of $\Gamma_1$
and $E_0$ cancel out each other. As a result, we simply adjust
$\Gamma_1$ to re-fit the spectrum until $HR^{mod}$ is consistent with
$HR^{obs}$ within the error range. These spectral parameters
($\Gamma_1$, $E_p$, $\Gamma_2$) are then taken to perform
extrapolation to estimate $S_\gamma$. Since the determinations of both
$\Gamma_1$ and $\Gamma_2$ are not fully based on fitting procedures,
it is very difficult to quantify their errors, and we only report
their estimated values. The error of $E_0$ is taken whenever possible
based on the best Band-function fit (see Table 1).

Case II: $\Gamma^{PL} > -1.2$ (2 out of 24 in the sample: GRBs 050726
and 050826).  In this case, $E_0$ should be far beyond
the BAT band, and BAT only covers the low energy part of the
spectrum ($E<E_0$). One has $\Gamma_1 \sim \Gamma^{PL}$ or slightly
larger (if $E_p$ is not very far above the band). It is very
difficult to estimate $E_0$ ($E_p$)
in this case. Nonetheless, one could use $HR^{obs}$ to pose some
constraints. Figure \ref{HR} suggests that $HR^{mod}$ converges to a
maximum value at high $E_p$'s given a certain $\Gamma_1$. We first let
$\Gamma_1=\Gamma^{PL}$ and check whether $HR^{obs}$ is consistent with
$HR^{mod}$. For both bursts in our sample one has $HR^{obs} >
HR^{mod}_{max}$. This suggests that $\Gamma_1$ should be larger than
$\Gamma^{PL}$. We then 
gradually increase $\Gamma_1$ until $HR^{mod}_{max}(\Gamma_1)$ becomes
consistent with $HR^{obs}$. Using this $\Gamma_1$ one can then fit for
$E_0$ (and $E_p$). According to Fig.\ref{HR}, there is great
degeneracy of $E_0$ ($E_p$) at $HR^{mod}_{max}$. The fitted
$E_0$ ($E_p$) therefore has very large errors and is unstable. In
practice, one could only set a lower limit of $E_0$ ($E_p$) below
which $HR^{mod}$ starts to deviate from $HR^{obs}$. As a result, only
lower limits of $E_p$ of these two bursts are reported in Table 1.

Case III: $\Gamma^{PL}<-2.3$ (6 out of 24 in the sample). In this
case, the burst is likely an XRF with $E_p$ near or below the low
energy end of BAT. BAT's observation likely only covers the high
energy part of the spectrum ($E>E_0$).
In order to constrain the spectral parameters using the hardness ratio
data, we assume $\Gamma_1=-1$ and $\Gamma_2=\Gamma^{PL}$ and fit for
$E_0$ by requiring $HR^{mod}(E_0)=HR^{obs}$. According to
Fig.\ref{HR}, near $HR^{mod}_{min}$, $HR^{mod}$ is rather insensitive
to $E_0$ ($E_p$). For most of the cases, one could find a solution of
$E_0$, but with very large errors. The solutions are also
unstable. For these cases, we do not report errors in Table 1, but
only report the best fit value with a $\sim$
symbol\footnote{Similarly, GRBs 050319, 050724, 050801 in the Case I
also have unstable solutions, so that the errors of their $E_p$'s are
not reported.}. For two cases (GRBs 050416A and 050819), there is no
solution since $HR^{obs}<HR^{mod}_{min}$. Similar to the Case II, we
then lower $\Gamma_2$ until a solution is found. Due to the degeneracy
of $E_0$ ($E_p$) with $HR$, only upper limits of $E_0$ ($E_p$) are
found, which are reported in Table 1.

\begin{figure*}
\epsscale{1.0} 
\plotone{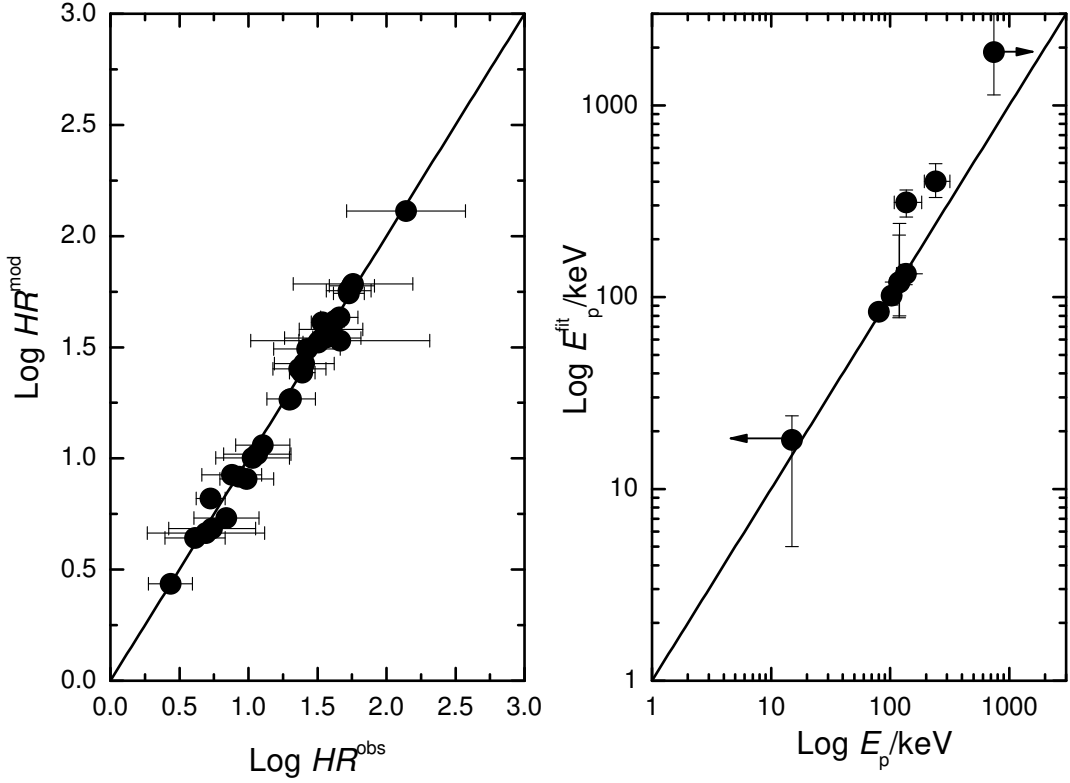} 
\caption{(a) Comparison of $HR^{mod}$ with $HR^{obs}$.  (b) Comparison
of $E_p$ derived from our method with $E_p^{fit}$ derived from the
joint fit using BAT, Konus-Wind, or HETE-2 data for eight bursts,
including GRBs 050128, 050219A, 050401, 050525A, 050713A, 050716,
050717, and 051221. The solid line is $E_p=E_p^{fit}$.}
\label{test}
\end{figure*}

With the spectral parameters derived from the above method, we have
extrapolated the observed 15-150 keV fluence $S_{\gamma,obs}$ to a
broader energy range ($1-10^4$ keV).  The derived $S_\gamma$ is
regarded as the total energy output during the prompt phase for
further efficiency studies. The results of prompt emission data are
reported in Table 1.  We caution that due to the intrinsic
instrumental limitation, the uncertainties of the results are
large. Nonetheless, we have made the best use of the available data
(especially the hardness ratio information) to derive the
parameters. Figure \ref{test} displays the robustness of the
method. Fig.\ref{test}a shows the criterion adopted in analyzing each
burst (eq.[\ref{criterion}]).  Fig.\ref{test}b shows how the
$E_p$'s derived with our method compares with the $E_p^{fit}$ derived
from the Konus-Wind or HETE-2 data for 8 bursts. The result suggests
that our derived $E_p$ meets $E_p^{fit}$ well for moderate $E_p$'s
(i.e. those falling into the BAT band), but deviate from $E_p^{fit}$
when $E_p$ is very large. Notice that the reported $E_p$ errors in our
method are derived from the best fit by fixing $\Gamma_1$ and
$\Gamma_2$. The real errors should also include the uncertainties of
$\Gamma_1$ and $\Gamma_2$. In such cases, the errors of $E_p$'s could
be larger to be more
consistent with $E_p^{fit}$ in the high-$E_p$ regime. However, due to
the difficulty of estimating these errors, they are not included in
reported errors.

\begin{figure*}
\epsscale{1.0} \plotone{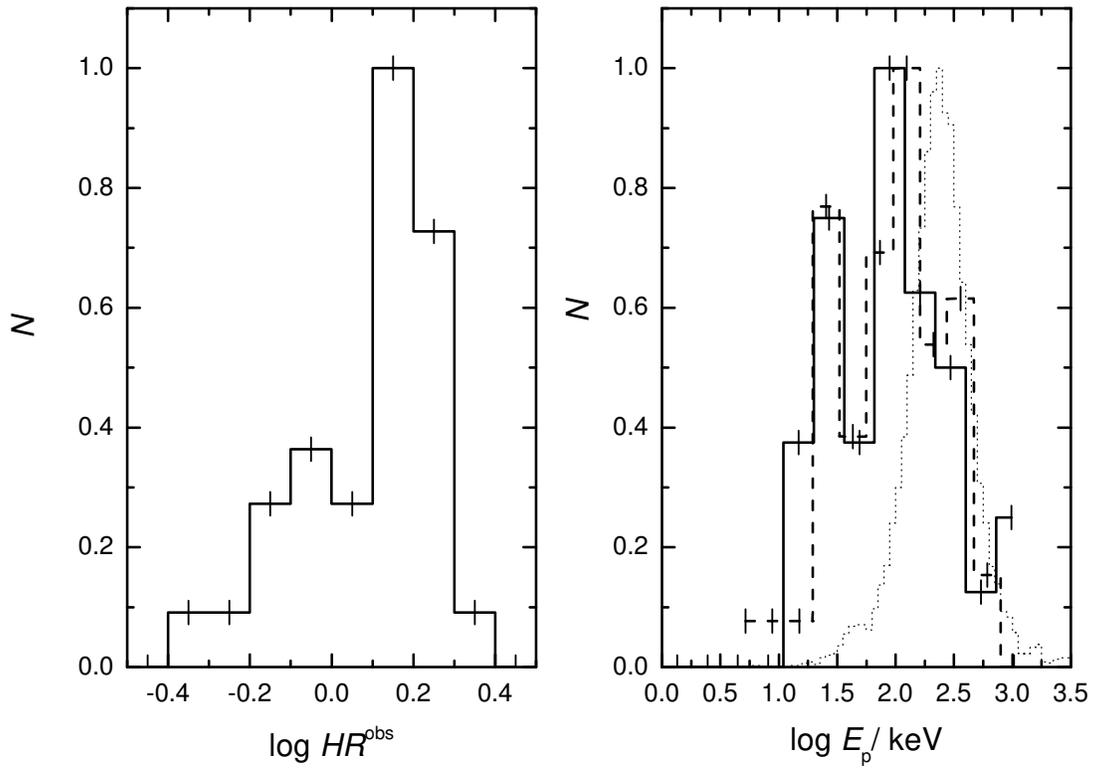}
\caption{The distributions of $HR^{obs}$ (left) and $E_p$ (right)
derived from our method. For the $E_p$ distribution, the histograms of
BATSE (dotted line) and HETE-2 (step-dashed line)
samples are also plotted.}
\label{dist1}
\end{figure*}

\begin{figure*}
\epsscale{1.0}
\plotone{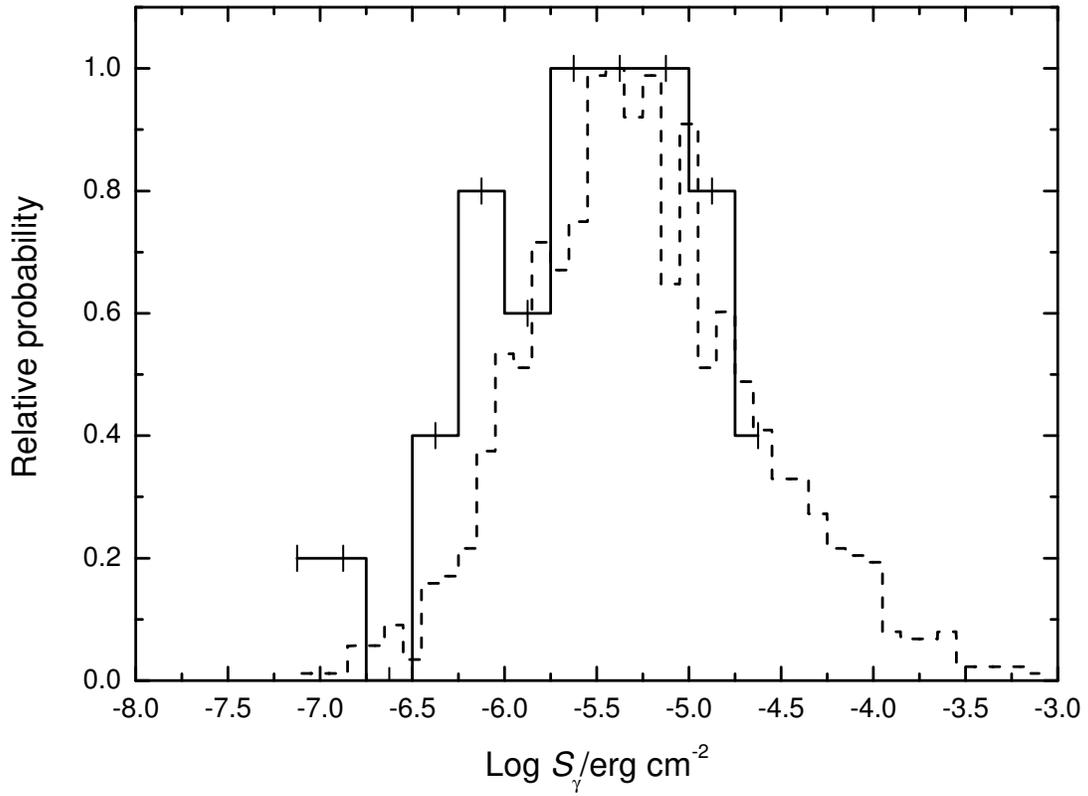}
\caption{Comparison of $\log S_{\gamma}$ distribution (in the 20-2000
keV band) of our sample (solid histogram) to that of the BATSE sample
(dashed histogram). }
\label{dist2}
\end{figure*}

In Fig.\ref{dist1}, we show the distributions of $HR$ and $E_p$ of our
sample. For the $E_p$ distribution (right panel), those of the BATSE
sample (Preece et al. 2000) and the HETE-2 sample (Lamb et al. 2005)
are also plotted for comparison. Our sample is generally consistent
with the HETE-2 sample, and tends to be softer than the BATSE sample
(as is expected because of a softer bandpass of BAT as compared with
BATSE). If one defines XRFs as those bursts with $HR^{obs}<1$, the number
ratio of the XRFs and the GRBs in our sample is $\sim 1:2$, being
consistent the HETE-2 result (Lamb et al. 2005). An interesting
feature is the marginal bimodal distribution of XRFs and GRBs, which
is consistent with the previous result derived with the HETE-2 data
(Liang \& Dai 2004, cf.  Sakamoto et al. 2005). In Fig.\ref{dist2}, we
plot the distribution of $S_\gamma$ as compared with the BATSE
results. The energy range of $S_\gamma$ is re-scaled to 20-2000 keV
(BATSE's energy band). We find 
that the two distributions are generally consistent with each other,
except that Swift GRB sample extends the BATSE sample to lower
fluences. This is expected because Swift is more sensitive than
BATSE.

\clearpage

%\clearpage

\clearpage

\subsection{X-ray afterglows}
\label{sec:XRT}

The X-ray afterglow light curves and the spectra of the bursts in our
sample have been presented by O'Brien et al. (2006). Among the 5
components of the synthetic X-ray lightcurve (e.g. Zhang et al. 2006),
the steep decay component is due to the GRB tail 
emission (Kumar \& Panaitescu 2000; Zhang et al. 2006) and the X-ray
flares are due to the late central engine activity (Burrows et al.
2005b; Zhang et al. 2006; Fan \& Wei 2005; Ioka et al. 2005; Liang et
al. 2006; Wu et al. 2006; King et al. 2005;
Perna et al. 2006; Proga \& Zhang 2006; Dai et al. 2006). We therefore
identify the steep decay component as well as the X-ray flares and
remove them from the light curve contribution.  We then fit the light
curves by either a broken power law for those bursts with
shallow-to-normal transition, or by a single power law
otherwise. Table 2 lists the X-ray data and the fitting results of
our sample.

\section{Prompt Gamma-ray fluence vs. X-ray Afterglow Fluence}
\label{sec:rel-eff}

In order to calculate the absolute value of the GRB radiative
efficiency $\eta_\gamma$ (eq.[\ref{eta}]), both $E_\gamma$ and $E_{K}$
need to be derived. This requires detailed modeling of $E_{K}$ and
the redshift information. The results depend on some unknown
parameters (e.g. the shock electron/magnetic field equipartition
factors, $\epsilon_e$, $\epsilon_B$, etc), which we discuss in detail
in \S\ref{sec:model}.  Nonetheless, using the directly measured
quantities listed in Tables 1 \& 2, one can analyze the relative
energetics between the prompt emission and the afterglow. The prompt
emission fluence $S_\gamma$ is a rough measure of the prompt emission
energetics. According to the standard afterglow model and assuming
that electron spectral index is $p \sim 2$, that the X-ray band is
above both the typical synchrotron emission frequency $\nu_m$ and the
synchrotron cooling frequency $\nu_c$, and that the inverse Compton
cooling is unimportant, the afterglow kinetic energy could be roughly
indicated by the quantity $\epsilon_e^{-1} F_{\nu,X}(t) t$, where $t$
is a particular epoch in the afterglow phase (e.g.  Freedman \& Waxman
2001; Berger et al. 2003; Lloyd-Ronning \& Zhang 2004). Since the
quantity $S_{X}(t) = F_{\nu,X}(t) t$ also has the dimension of
fluence, the $S_\gamma$-to-$S_{X}(t)$ ratio could give a rough
indication of the relative energetics between the prompt emission and
the afterglow. The unknown redshift essentially does not enter the
problem.

The shallow decay phase commonly observed in {\em Swift} GRBs has been
generally interpreted as a refreshed external shock (e.g. Rees \&
M\'esz\'aros 1998; Dai \& Lu 1998; Panaitescu et al.  1998; Kumar \&
Piran 2000; Sari \& M\'esz\'aros 2000; Zhang \& M\'esz\'aros 2001,
2002a; Dai 2004; Zhang et al. 2006; Nousek et al. 2006; Panaitescu et
al. 2006; Granot \& Kumar 2006). Within this interpretation, the
kinetic energy of the afterglow $E_K$ increases with time for an
extended period.  This brings extra complication to the efficiency
problem. One needs to identify at which epoch the corresponding $E_K$
represents the kinetic energy left over right after the prompt
gamma-ray emission. This is a model-dependent problem, and we take an
approach to accommodate different possibilities. We pay special
attention to two epochs. One is the break time $t_b$ at the
shallow-to-normal-decay transition epoch, which corresponds to the
epoch when the putative injection phase is cover. Within the injection
interpretation, another important time is the fireball deceleration
time ($t_{\rm dec}$), which is usually earlier or around the
first data point in the shallow decay phase. The kinetic energy at
$t_{\rm dec}$ [$E_{K}(t_{\rm dec})$] is relevant to the GRB efficiency
problem, if the injected energy during the shallow decay phase is due
to a long-term central engine (e.g.  Dai \& Lu 1998; Zhang \&
M\'esz\'aros 2001; Dai 2004), since the bulk of kinetic energy is
injected at later epochs. In the scenario that the injection is due to
an instantaneous injection with variable Lorentz factors (Rees \&
M\'esz\'aros 1998; Kumar \& Piran 2000; Sari \& M\'esz\'aros 2000;
Zhang \& M\'esz\'aros 2002a; Granot \& Kumar 2006), the total kinetic
energy of the outflow right after the prompt emission is over should
be defined by the kinetic energy measured at $t_b$ [$E_{K}(t_b)$] when
the injection phase is over. However, the kinetic energy of {\em the
ejecta} that gives rise 
to the gamma-ray emission may be still roughly $E_K(t_{\rm dec})$,
since the gamma-ray emission from the low-Lorentz-factor ejecta can
not escape due to the well-known compactness problem (e.g.  Piran
1999). Nonetheless, some other scenarios do not interpret the shallow
decay as additional energy injection (e.g. Eichler \& Granot 2006;
Toma et al. 2006; Ioka et al.  2006; Kobayashi \& Zhang 2006). In some
of these cases, $E_{K}(t_b)$ is the more relevant quantity to define
the GRB efficiency, e.g. in the models invoking precursors (e.g. Ioka
et al. 2006) and the models involving delayed energy transfer
(e.g. Kobayashi \& Zhang 2006; Zhang \& Kobayashi 2005). In this paper
we use both $E_{K}(t_{\rm dec})$ and $E_{K}(t_b)$ to define GRB
radiative efficiencies.

The injection break time $t_b$ can be directly measured from the light
curves. The fireball deceleration time (for burst durations shorter
than this time scale, the so-called thin shell regime) $t_{\rm
dec}\sim 5(1+z)(E_{K,52}/n)^{1/3}(\gamma_0/300)^{-8/3}$, on the other
hand, is not directly measured and is very likely buried
beneath the steep-decay prompt emission tail component. Here $E_{K,52}
= E_K/10^{52} ~{\rm ergs}$ (the convention $Q_n=Q/10^n$ in cgs
units is adopted throughout the paper), $n$ is the ambient medium
density in unit of 1 proton / cm$^3$, and $\gamma_0$ is the initial
Lorentz factor of the fireball. Without knowing $\gamma_0$ (noticing
the sensitive dependence on this unknown parameter), one cannot
accurately estimate this time with the observables. For
typical parameters (e.g.  $z=1$, $\gamma_0=150$, and $E_{K,52}/n=1$)
we get $t_{\rm dec} \sim 60$s.  Considering also the thick shell
regime (i.e. the burst duration is longer than the above critical
time, Kobayashi et al. 1999), we finally roughly estimate the
deceleration time as $t_{\rm dec}\sim {\rm max} (60~{\rm s}, T_{90})$.

\begin{figure*}
\epsscale{.8} 
\plotone{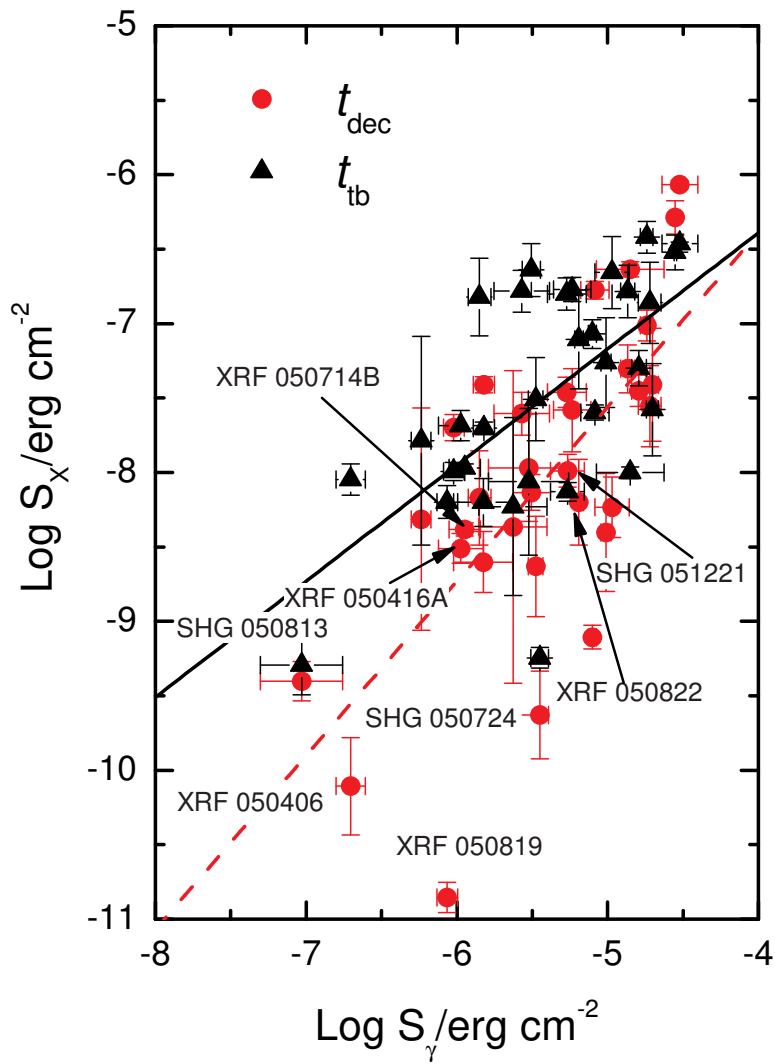} 
\caption{The extrapolated prompt emission fluence $S_\gamma$ against
the X-ray fluence at $t_{\rm dec}$ (dots) and at $t_b$ (triangles).
The solid and dashed lines, with slopes of 0.78 and 1.17, are the best
fit to the data at $t_b$ and $t_{dec}$, respectively. The short-hard
GRBs (SHGs) and X-ray flashes (XRFs) are marked.} 
\label{Sg_Sx}
\end{figure*}

In Figure \ref{Sg_Sx} we plot $S_X(t_{\rm dec})$ (red dots) and
$S_X(t_b)$ (black triangles) against $S_\gamma$. The large differences
between $S_{X}(t_{\rm dec})$ and $S_{X}(t_b)$ indicate that
significant energy injection happens in many bursts. One interesting
signature evidenced in Fig.\ref{Sg_Sx} is that $S_\gamma$ is
positively correlated to $S_X$. This is consistent with the previous
knowledge that the radiated energy is positively correlated to the
afterglow kinetic energy. For the case of $S_X(t_b)$, the
$S_\gamma-S_X$ correlation slope is $0.78\pm 0.17$, which is shallower
than unity. This suggests that considering the total kinetic energy
when the energy injection phase is over [$E_{K} (t_b)$], the fainter
bursts are not as efficient as brighter ones in converting kinetic
energy into radiation. This is consistent with the pre-{\em Swift}
finding of a shallow correlation between the efficiency $\eta_\gamma$
and $E_\gamma$ (Lloyd-Ronning \& Zhang 2004; Lamb et al. 2005). Since
generally there is a positive correlation between the $E_\gamma$ and
$E_p$ (Lloyd et al. 2000; Amati et al.  2002; Liang et al. 2004a; Lamb
et al. 2005), this also suggests that softer bursts (e.g.  XRFs) tend
to have lower efficiencies (Soderberg et al. 2004; Lloyd-Ronning \&
Zhang 2004). For the case of $t_{\rm dec}$, however, $S_\gamma - S_X$
correlation slope is very close to unity ($1.17\pm 0.22$).  This means
that the efficiency defined by $E_K(t_{dec})$ (the ``true'' efficiency
in the models invoking additional energy injection) is not sensitively
related to $E_\gamma$, and hence, to $E_p$ according to the positive
$E_\gamma-E_p$ correlations.  This means that XRFs are not
intrinsically inefficient GRBs if the injection hypothesis is true.

\begin{figure*}
\epsscale{1.0} 
\plotone{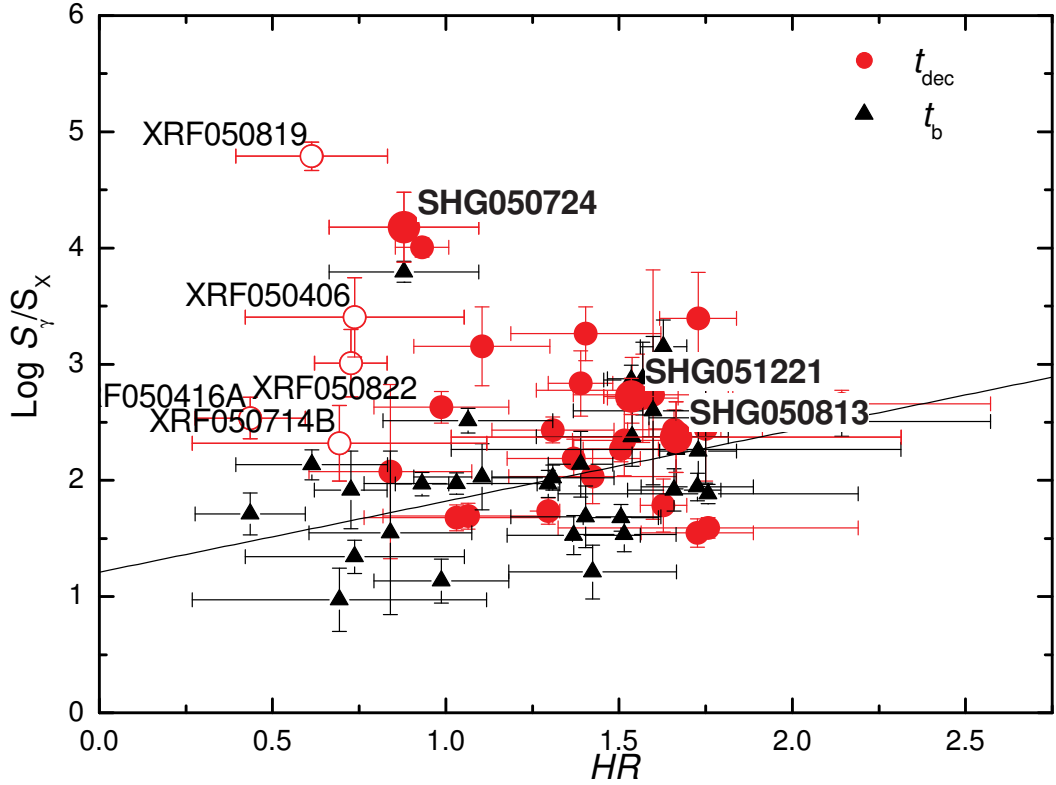} 
\caption{Ratios of $S_{\gamma}/S_X$ against the hardness ratio
$HR$. The circles represent the data at $t_{dec}$, and the triangles
represent the data at $t_b$. The short-hard GRBs (larger solid
circles) and the XRFs (open circles) at $t_{dec}$ are marked. The
straight line is the best fit of the data at $t_b$, which shows a
shallow correction between the two quantities, with a linear
correlation coefficient 0.61 and a chance probability $<0.01$.}
\label{Ep_R}
\end{figure*}

To show the effect more clearly, we plot the ratios $R(t_{\rm
dec})=S_{\gamma}/S_X (t_{\rm dec})$ (red dots) and
$R(t_{t_b})=S_{\gamma}/S_{X}(t_{b})$ (black triangles) against the
hardness ratio $HR^{\rm obs}$ (Fig.\ref{Ep_R}). It is found that while
the soft bursts (e.g. XRFs) tend to have a lower gamma-to-X ratio (and
hence, lower radiative efficiency) than the typical GRBs at $t_b$ (a
shallow linear dependence with index $0.61\pm 0.21$,
they do not differ too much from hard GRBs at $t_{\rm dec}$.  This
means that XRFs are radiatively as efficient as hard GRBs. Such a
possibility has been speculated by Lloyd-Ronning \& Zhang
(2004) and was first recognized by Schady et al. (2006) when analyzing
the early UVOT data of XRF 050406. Now we extend the analysis to a
larger sample and verify that it is common for other XRFs as well. Such
a conclusion is strengthened by a more careful treatment of radiative
efficiency in \S\ref{sec:eta}, and we discuss the implications of this
result for the XRF models in \S5.

\clearpage
\clearpage

\clearpage

%%%%%%%%%%%%%%%%%%%%%%%%%%%%%%%%%%%%%%%%%%%%%%%%%%%%%%%%%%%%%%%%%%%%
\section{GRB efficiency}
\label{sec:model}

\subsection{Theoretical Models}
\label{sec:EK}

In this section we explicitly derive the radiative efficiency for the
GRBs in our sample according to eq.(\ref{eta}). The isotropic prompt
emission energy $E_\gamma$ is derived from the extrapolated 1-10000
keV band fluence ($S_\gamma$) according to
\begin{eqnarray}
E_\gamma & = & {4 \pi D_L^2} S_\gamma {(1+z)^{-1}}\nonumber 
\\ & = & 1.3\times 10^{51}~{\rm erg}~ D_{L,28}^2(1+z)^{-1} S_{\gamma,-6} 
\label{Egam}
\end{eqnarray}
where $D_L$ is the luminosity distance.

The derivation of $E_K$ requires detailed afterglow
modeling. Regardless of whether there is indeed a long-lasting
central engine, the energy injection process could be mimicked by
introducing an ``effective'' long lasting central engine with
luminosity $L=L_0 (t/t_0)^{-q}$. For the varying Lorentz factor
injection model the Lorentz factor index $s$ could be related to an
effective $q$ through $q=(10-2s)/(7+s)$ for an ISM medium and
$q=4/(3+s)$ for a wind medium (Zhang et al. 2006). The
injection then results in an evolving kinetic energy $E_K \propto
t^{(1-q)}$. After the energy injection is over, the fireball can be
described by the standard afterglow model. At any epoch during the
injection phase, the afterglow emission level could be calculated by
taking the kinetic energy at that time. So in our treatment, we still
use the standard afterglow model to derive various parameters as
functions of $E_K$, bearing in mind that $E_K$ may be time-dependent.

For a constant density medium, the typical synchrotron emission
frequency, the cooling frequency and the peak spectral flux read (Sari
et al. 1998; coefficients taken from Yost et
al. 2003)\footnote{Lloyd-Ronning \& Zhang (2004) adopted the
coefficients from Hurley et al. (2002). The $\nu_m$ coefficient is
larger than adopted here. This effect, together with the ignorance of
the Inverse Compton (IC) effect, leads to systematic underestimating
of $E_K$ and overestimating of $\eta_\gamma$, as also pointed out by
Fan \& Piran (2006) and Granot et al. (2006). This systematic
deviation does not affect the global dependences of $\eta_\gamma$
on other parameters, as are reproduced in this paper.}

\begin{eqnarray}
\nu_m & = & 3.3 \times 10^{12} {\rm\ Hz}
\left(\frac{p-2}{p-1}\right)^2(1+z)^{1/2}\epsilon_{B,-2}^{1/2}
\epsilon_{e,-1}^{2}E_{K,52}^{1/2} t_d^{-3/2} \label{num}\\
\nu_c & = & 6.3 \times 10^{15} {\rm\ Hz}
(1+z)^{-1/2} (1+Y)^{-2} \epsilon_{B,-2}^{-3/2}E_{K,52}^{-1/2} n^{-1}
t_d^{-1/2} \label{nuc}\\ F_{\nu,\max} & = & 1.6 {\rm\ mJy}
(1+z)D^{-2}_{28}\epsilon_{B,-2}^{1/2}E_{K,52}n^{1/2}~
\label{Fnumax}.
\end{eqnarray}
Here $t_d$ is the observer's time in unit of days,
\be
Y=[-1+(1+4 \eta_1\eta_2 \epsilon_e / \epsilon_B)^{1/2}]/2
\label{Y}
\ee
is the IC parameter, where $\eta_1 = {\rm min} [1,(\nu_c/
\nu_m)^{(2-p)/2}]$  (Sari \& Esin 2001),
and $\eta_2 \leq 1$ is a correction factor introduced by the
Klein-Nishina correction. The latter effect was treated in detail by
Fan \& Piran (2006). Here we adopt an alternative, approximate
treatment. For the electron Lorentz factor $\gamma_{e,X}$
corresponding to the X-ray band emission, the synchrotron self-IC
effect is significantly suppressed in the Klein-Nishina regime for the
photons with energy $\nu > \nu_{\rm KN}$, where
\begin{eqnarray}
\nu_{\rm KN} & = & h^{-1} \Gamma m_e c^2 \gamma_{e,X}^{-1} (1+z)^{-1}
\nonumber \\
& \simeq & 2.4 \times 10^{15} ~{\rm Hz}~ (1+z)^{-3/4} E_{52}^{1/4}
\epsilon_{B,-2}^{1/4} t_d^{-3/4} \nu_{18}^{-1/2} ~,
\end{eqnarray}
and $h$ is the Planck's constant. Based on the $\nu F_\nu$ spectrum
of the standard synchrotron emission
model (Sari et al. 1998), one can roughly estimate $\eta_2 = {\rm min}
[1, (\nu_{\rm KN} / \nu_c)^{(3-p)/2}]$ for slow cooling ($\nu_m <
\nu_c$) and $\eta_2 = {\rm min} [1, (\nu_{\rm KN} / \nu_m)^{1/2}]$ for
fast cooling ($\nu_c < \nu_m$), where the factors $(\nu_{\rm KN} /
\nu_c)^{(3-p)/2}$ and $(\nu_{\rm KN} / \nu_m)^{1/2}$ denote the
fractions of the photon energy density that contributes to self-IC in
the X-ray band in the slow and fast cooling regimes, respectively.

In previous analyses (e.g. Freedman \& Waxman 2001; Berger et
al. 2003; Lloyd-Ronning \& Zhang 2004), the IC cooling was usually not
taken into account.  The inclusion of IC cooling modifies the X-ray
afterglow light curves considerably (e.g. Wu et al. 2005), which also
influences the derived GRB efficiency $\eta_\gamma$. Here we generally
include the IC factor [power of $(1+Y)$] in the treatment (see also
Fan \& Piran 2006). The results are reduced to the previous pure
synchrotron-dominated case when $Y \ll 1$.

For about 2/3 cases in our sample (22 out of 31), the X-ray band
temporal decay index 
and the spectral index are consistent with the spectral regime $\nu >
{\rm max} (\nu_m, \nu_c)$. This is the regime where $E_K$ is
independent of $n$ and only weakly depends on $\epsilon_B$ and $p$,
and therefore an ideal regime to measure $E_K$. One can derive the
X-ray band energy flux as\footnote{Although our treatment is for a
constant-density medium, eq.(\ref{FnuX-1}) is valid for more general
cases (e.g. wind medium) since in this regime the flux does not depend
on the medium density.}
\begin{eqnarray}
\nu F_{\nu} (\nu=10^{18}~{\rm Hz})& = &
F_{\nu,\max}\nu_c^{1/2}\nu_m^{(p-1)/2}\nu_X^{(2-p)/2}
\nonumber \\
& = & 5.2\times 10^{-14} ~{\rm ergs~s^{-1} ~cm^{-2}}
D_{28}^{-2}(1+z)^{(p+2)/4} \nonumber \\ & \times & (1+Y)^{-1} f_p
\epsilon_{B,-2}^{(p-2)/4} \epsilon_{e,-1}^{p-1} E_{K,52}^{(p+2)/4}
t_d^{(2-3p)/4}\nu_{18}{^{(2-p)/2}}~,
\label{FnuX-1}
\end{eqnarray}
where
\be
f_p = 6.73 \left(\frac{p-2}{p-1}\right)^{(p-1)} (3.3\times
10^{-6})^{(p-2.3)/2}
\ee
is a function of $p$, which is calculated in Figure 7. It peaks at
$\sim 1.72$ when
$p \sim 2.12$, and declines monotonically at large $p$ values. For
example, at $p \sim 3$, $f_p$ is only $\sim 0.02$. This gives a nearly
2 orders of magnitude variation for $p=(2.01 - 3)$, and thus demands
more careful treatments of individual bursts presumably having quite
different $p$ values.

With eq.(\ref{FnuX-1}), one can derive $E_K$ at any time $t_d$ as
\begin{eqnarray}
E_{K,52} & = & \left[\frac{\nu F_\nu (\nu=10^{18}~{\rm Hz})}{5.2\times
10^{-14} ~{\rm ergs~s^{-1} ~cm^{-2}} }\right]^{4/(p+2)}
D_{28}^{8/(p+2)}(1+z)^{-1} t_d^{(3p-2)/(p+2)}
\nonumber \\
& \times & (1+Y)^{4/(p+2)} f_p^{-4/(p+2)}
\epsilon_{B,-2}^{(2-p)/(p+2)}
\epsilon_{e,-1}^{4(1-p)/(p+2)}
\nu_{18}{^{2(p-2)/(p+2)}}
\label{EK1}
\end{eqnarray}
This could be reduced to $E_K \sim  \epsilon_e^{-1}S_X D_L^2 / (1+z)$
for $p \sim 2$. Comparing with eq.(\ref{Egam}), we can see that as far
as the efficiency problem is concerned the redshift-dependence is very
weak.
\clearpage
\begin{figure*}
\epsscale{1.0}
\plotone{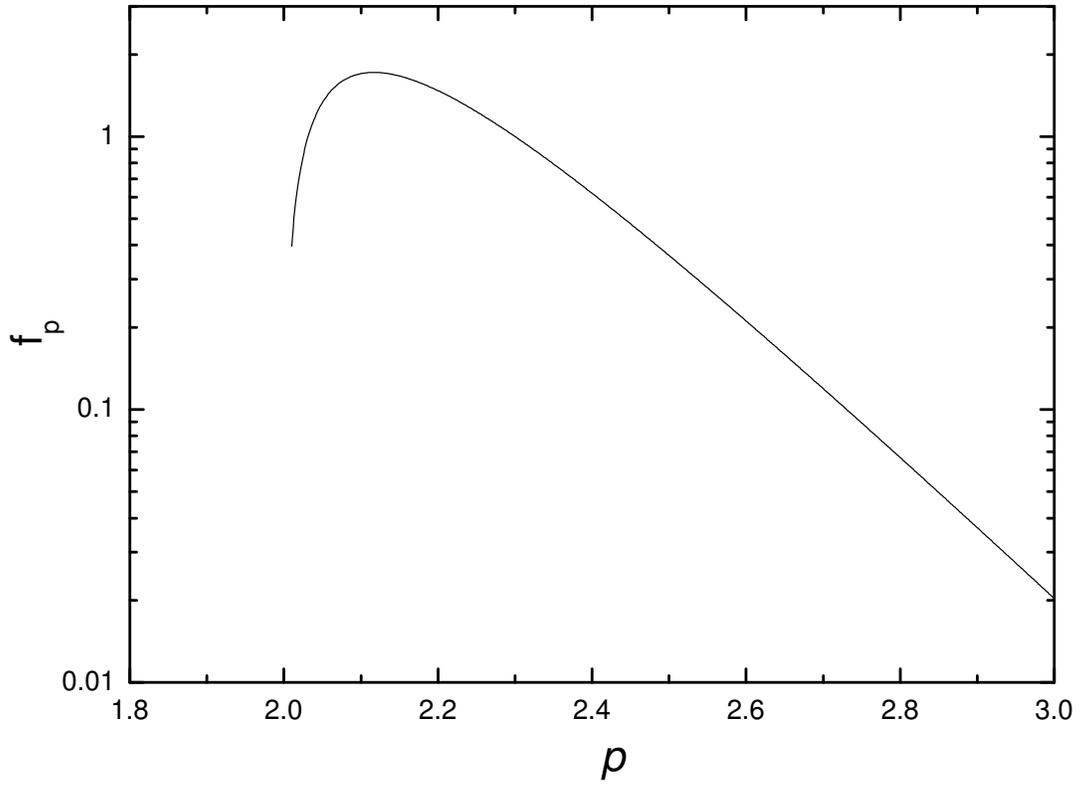}
\caption{The function $f_p$. }
\end{figure*}
\clearpage

For nearly 1/3 cases in our sample (9 out of 31), the X-ray data are
not consistent 
with being in the regime $\nu_X > {\rm max} (\nu_m, \nu_c)$ (which
requires $\alpha = (3 \beta -1)/2$ for the $F_\nu \propto t^{-\alpha}
\nu^{-\beta}$ convention). The temporal decay slope in the normal
decay phase is close to -1. This rules out the fast cooling case
$\nu_c < \nu_X < \nu_m$ for both ISM and wind models. For slow cooling
models ($\nu_m < \nu_X < \nu_c$), the temporal decay index derived
from the wind model [$\alpha = (3\beta +1)/2$ for the $F_\nu \propto
t^{-\alpha} \nu^{-\beta}$ convention, Chevalier \& Li (2000)] is too
steep compared with the data in our sample. One is then left with the
only possibility, $\nu_m < \nu_X < \nu_c$ in the ISM model. In fact
the observed $\alpha$ and $\beta$ are consistent with being in this
regime [$\alpha = (3/2)\beta$]. The derived X-ray
band energy flux is then
\begin{eqnarray}
\nu F_{\nu} (\nu=10^{18}~{\rm Hz})& = &
F_{\nu,\max} (\nu_m/\nu_X)^{(p-1)/2}
\nonumber \\
& = & 6.5\times 10^{-13} ~{\rm ergs~s^{-1} ~cm^{-2}}
D_{28}^{-2}(1+z)^{(p+3)/4} \nonumber \\ & \times & f_p
\epsilon_{B,-2}^{(p+1)/4} \epsilon_{e,-1}^{p-1} E_{K,52}^{(p+3)/4}
n^{1/2} t_d^{(3-3p)/4}\nu_{18}{^{(3-p)/2}}~,
\label{FnuX-2}
\end{eqnarray}
This gives
\begin{eqnarray}
E_{K,52} & = & \left[\frac{\nu F_\nu (\nu=10^{18}~{\rm Hz})}{6.5\times
10^{-13} ~{\rm ergs~s^{-1} ~cm^{-2}} }\right]^{4/(p+3)}
D_{28}^{8/(p+3)}(1+z)^{-1} t_d^{3(p-1)/(p+3)}
\nonumber \\
& \times & f_p^{-4/(p+3)} \epsilon_{B,-2}^{-(p+1)/(p+3)}
\epsilon_{e,-1}^{4(1-p)/(p+3)} n^{-2/(p+3)}
\nu_{18}{^{2(p-3)/(p+3)}}
\label{EK2}
\end{eqnarray}
Inspecting eq.(\ref{nuc}), one can draw the conclusion that in order
to have $\nu_c > \nu_X$ in the normal decay regime ($t_d \sim 1$),
$\epsilon_B$ must be very small (e.g. $< (10^{-3}- 10^{-4})$). Keeping
a more or less constant $\epsilon_e \sim 0.1$, the $Y$ parameter
(eq.[\ref{Y}]) does not increase significantly for a smaller
$\epsilon_B$, since the $\eta_2$ parameter becomes much smaller due to
the Klein-Nishina suppression. In order not to derive a unreasonably
large $E_K$ value (limited by the total energy budget of the
progenitor system), the data also require a small ambient density
(e.g. $n<0.1$).

\subsection{Calculation Results}
\label{sec:eta}

Eqs. (\ref{EK1}) and (\ref{EK2}) suggest that the absolute value of
$E_{K}$ depends on several unknown shock parameters. In order to
calculate $\eta_\gamma$, the absolute value of $E_{K}$ is needed. This
requires a detailed multi-wavelength study (e.g. Panaitescu \& Kumar
2002; Yost et al. 2003) to constrain unknown shock parameters as
well. Limited by the X-ray data alone, one inevitably needs to make
some assumptions on the unknown shock parameters.

Our first step is to use the X-ray data (temporal index $\alpha_X$ and
spectral index $\beta_X$) to determine the spectral regime the burst
belongs to. In all the cases, we choose the ``normal'' decay phase for
the temporal index ($\alpha_{2}$ for the broken power-law fit, or
$\alpha_{1}$ for the single power-law fit if $\alpha_{1}$ is close to
unity). This is because this segment has no contamination of
energy injection (with unknown $q$ parameter). Using $\alpha_X$ and
$\beta_X$ we check the spectral regime of the X-ray band by comparing
the $\alpha-\beta$ relation of various models (e.g. Table 1 of Zhang
\& M\'esz\'aros 2004). We find that within the error uncertainties of
the data, most bursts could be grouped into two spectral regimes: (I)
$\nu_X > {\rm max} (\nu_m, \nu_c)$ (20 out of 31); and (II) $\nu_m <
\nu_X < \nu_c$ in the ISM model (9 out of 31).
Two bursts have unexpectedly large spectral indices, i.e. GRBs
050319 ($\beta_X=2.02\pm 0.47$) and 050714B ($\beta_X=4.50\pm
0.70$). The spectrum is likely dominated by the contribution of the
GRB tail emission and we assume that in the afterglow phase they are
in the regime of $\nu_X > {\rm max} (\nu_m, \nu_c)$, the default
case. GRBs 050215B ($\beta_X\sim 0.5\pm 0.5$) and 050716
($\beta_X=0.33\pm 0.03$) have very small spectral indices, and we
assume that it is in the regime $\nu_m < \nu_X < \nu_c$. After
determining the spectral regimes, we derive $p$ from $\beta_X$
($\beta_X =-p/2$ for $\nu_X > {\rm max} (\nu_m, \nu_c)$ and $\beta_X=
-(p-1)/2$ for $\nu_m < \nu_X < \nu_c$). Given the large error bars
usually associated with the spectral indices,
whenever $p\leq 2$ and $p>3$ we take $p=2.01$ and $p=3$, respectively.

\subsubsection{$\nu_X > {\rm max}(\nu_m, \nu_c)$}

Since there are too many unknown parameters for the regime II bursts
(eq.[\ref{EK2}]), we first ignore them and focus on the regime I
bursts, whose $E_K$ essentially only depend on $\epsilon_e$. Previous
broadband fitting suggests that $\epsilon_e$ is typically around 0.1
(Wijers \& Galama 1999; Panaitescu \& Kumar 2002; Yost et al. 2003;
Liang et al. 2004; Wu et al. 2004). The value of $\epsilon_B$ has a
large scatter for previous bursts but nonetheless has a typical value
of $0.01$ (e.g. Panaitescu \& Kumar 2002). The existence of regime II
bursts suggest that at least some bursts have very small
$\epsilon_B$. We therefore also consider the cases with a smaller
$\epsilon_B$, say, $\sim 10^{-4}$. In any case $E_K$ is insensitive to
$\epsilon_B$ in regime I.

Our calculation procedure for regime I GRBs is as follows.

1. Use the extrapolated $S_\gamma$ in the 1-10000 keV band to derive
$E_\gamma$ according to eq.(\ref{Egam}). Since $\eta_\gamma$ is
insensitive to $z$, we take a moderate redshift, $z=2$, for those
bursts whose redshifts are not directly measured.

2. Use $\beta_X$ and the 0.3-10 keV band flux to calculate the
monochromatic flux at $10^{18}$ Hz, $F_{\nu}(\nu = 10^{18}~{\rm Hz})$.

3. Calculate $E_{K,52}$ with eq.(\ref{EK1}) at two epochs, $t_{\rm
dec}$ and $t_b$. For each epoch we calculate two values: one value
($E_{K,52}^{(1)}$) for $(\epsilon_e,\epsilon_B)=(0.1,0.01)$, and
another ($E_{K,52}^{(2)}$) for $(\epsilon_e,
\epsilon_B)=(0.1,10^{-4})$. The $Y$ parameter is searched
self-consistently according to
the method described in \S\ref{sec:EK}.

4. Use eq.(\ref{eta}) to derive ${\eta}_\gamma^{(1)}$ and
${\eta}_\gamma^{(2)}$ at $t_{\rm dec}$ and $t_b$.

\begin{figure*}
\epsscale{1.0} \plotone{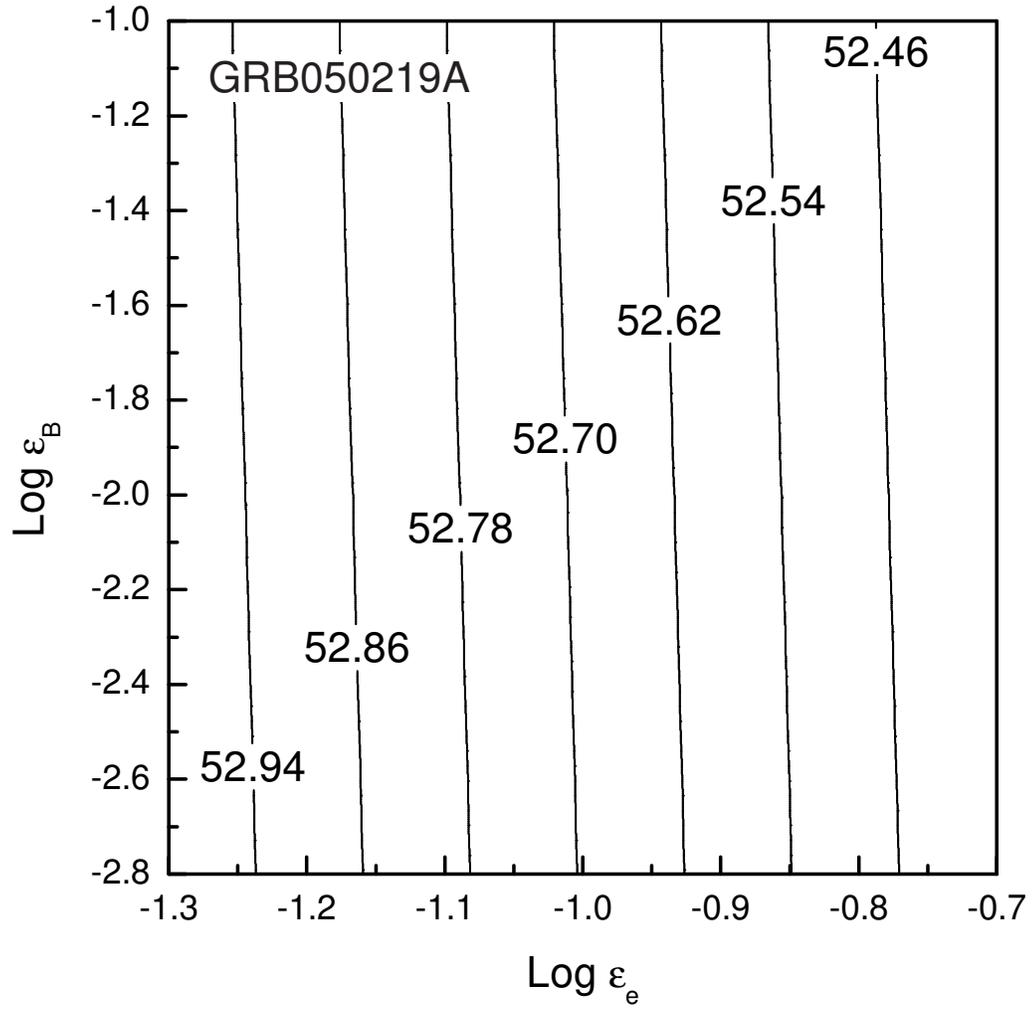}
\caption{Contours of $\log E_K$ in the $(\epsilon_e,
\epsilon_B)$ plane for GRB 050219A.} \label{contour}
\end{figure*}

Table 3 displays our calculation results for regime I GRBs. Equation
(\ref{EK1}) indicates that the apparent dependence on
$\epsilon_B$ is weak. This is strengthened by the Klein-Nishina effect
since $Y$ does not increase significantly as $\epsilon_B$ is
lowered. Comparing $E_{K}^{1}$ and $E_K^2$ (or $\eta_\gamma^1$ and
$\eta_\gamma^2$) at a same epoch, we can see that the difference
introduced by changing $\epsilon_B = 0.01$ to $10^{-4}$ is not
significant. In Fig.\ref{contour} we show the $E_K$ contour in the
$(\epsilon_e, \epsilon_B)$ plane for GRB 050219A. It again shows the
result is insensitive to $\epsilon_B$, so that $\epsilon_e$ is the
most sensitive parameter for determining $E_K$ and hence, for
determining $\eta_\gamma$. The existence of the regime II
bursts (which requires low values of $\epsilon_B$) suggests that if
shock parameters are not too different from burst to burst, the
$\epsilon_B$ value for the regime I bursts may be
also low (e.g. the second parameter set, $(\epsilon_e,
\epsilon_B)=(0.1,10^{-4})$, for the regime I calculations). This
suggests slightly lower radiative efficiencies than previously
estimated (typically taken as $\epsilon_B \sim 0.01$), since a lower
$\epsilon_B$ nonetheless slightly increases $E_K$ despite a very
shallow dependence.
\clearpage
\clearpage

\begin{figure*}
\epsscale{1.0}
\plotone{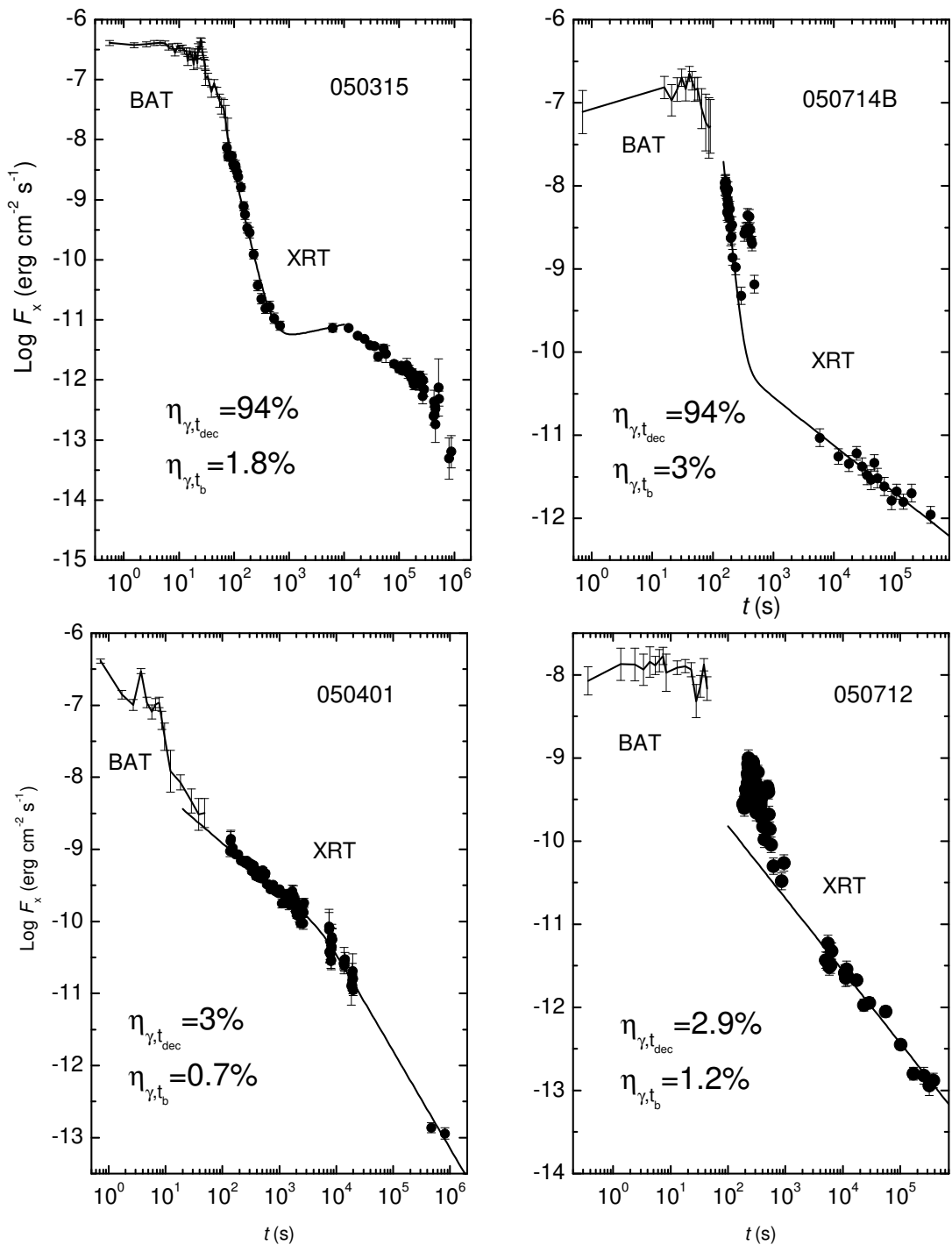}
\caption{Comparisons of the joint-BAT-XRT light curves in the XRT band
for some GRBs having extremely high or low efficiencies at $t_{dec}$.}
\label{lc}
\end{figure*}

Inspecting the calculated $\eta_\gamma$ for the bursts in the spectral
regime I, we find that at $t_{\rm dec}$, 10 out of 22 bursts have a
radiative efficiency $\eta_\gamma^{(1)}$ higher than $60\%$, sometimes
even as high as $98\%$ (GRB 050819).  The rest have much lower
efficiencies, sometimes only a few percent. At $t_{b}$, on the other
hand, the values of $\eta_\gamma$ are typically several percent or
even lower.  Those bursts with low efficiencies from the very
beginning correspond to the cases without significant energy injection
in the early phase. For illustration, in Fig.\ref{lc} we present the
combined BAT-XRT light curves in the XRT band for some GRBs having
extremely high or low efficiencies at $t_{\rm dec}$ (see also O'Brien
et al.  2006). It is evident that the XRT light curves of the
high-$\eta_{\gamma}(t_{dec})$ GRBs have a very flat energy injection
component and a prominent steeply decaying prompt emission tail
(e.g. GRB 050315 and GRB 050714B). Those with
low-$\eta_\gamma(t_b)$, on the other hand, typically have a smooth
transition from prompt emission to afterglow without a significant
steep decay component and/or a shallow decay component due to energy
injection (e.g. GRB 050401 and GRB 050712).
\clearpage
\clearpage

\begin{figure*}
\epsscale{1.0} 
\plotone{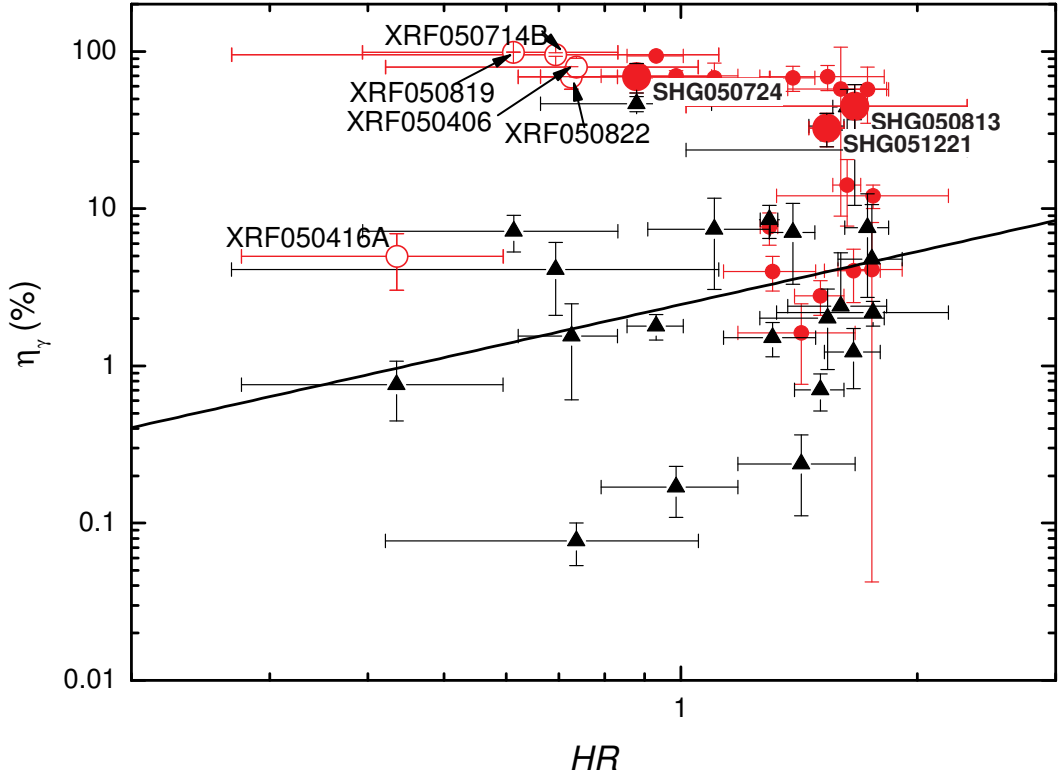} 
\caption{The radiative efficiency $\eta_\gamma^{(1)}$ against the
hardness ratio $HR$. The circles represent the data at $t_{dec}$, and
the triangles represent the data at $t_b$. The short-hard GRBs (larger
solid circles) and the XRFs (open circles) at $t_{dec}$ are also
marked. The dashed line is the best fit of the data at $t_b$, which
shows a shallow correction between the two quantities similar to that
shown in Figure 6.}
\label{eta-Ep}
\end{figure*}

In Fig.\ref{eta-Ep} we plot $\eta_\gamma^{(1)}$ as a function of
the hardness ratio $HR$. We again find that while there exists a
shallow correlation between $\eta_{\gamma,t_{b}}^{(1)}$ and $HR$ (with
index $1.12\pm 0.90$), such
a correlation essentially disappears when $\eta_\gamma$ at $t_{\rm
dec}$ is considered.  This is consistent with the analysis in
\S2.2. The results suggest that if $E_K(t_{dec})$ is the relevant
afterglow energy left over after the prompt emission, XRFs and GRBs
intrinsically have similar radiative efficiencies, as opposed to the
conclusion drawn using late-time X-ray data only.
\clearpage
\clearpage

Two short-hard GRBs 050724 and 050813 are in the spectral regime
I. The results in Table 3 suggest that there is no noticeable
difference between short, hard GRBs and long, soft GRBs as far as the
radiative efficiencies are concerned. The same conclusion was drawn
for the first short GRB with X-ray afterglow detection (GRB 050509b,
Gehrels et al. 2005; Bloom et al. 2006), and is being
verified by a larger sample of short GRBs in accumulation. 
To increase the short GRB sample, we have also performed the
same analysis to GRB 051221A (e.g. A. Parsons et al. 2006, in
preparation). The results are included in all the Tables and figures,
making the total number of short bursts in our sample to 32. The
inclusion of this burst strengthens the conclusion that short GRBs are
no different from long GRBs in radiative efficiencies.

\subsubsection{$\nu_m < \nu_X < \nu_c$}

For the regime II bursts, we can set up
an upper limit on $\epsilon_B$ using the light curves. Typically this
limit is very low (say, $<10^{-4}$) since the afterglow light curves
usually extend to very late epochs (say, $>1$ day). The efficiency
estimated for these bursts have even larger uncertainties because
$E_K$ depends on many unknown parameters, e.g. $\epsilon_e$,
$\epsilon_B$, $n$, etc.  Nonetheless, we perform some rough estimates
with the following procedure.

To keep $\nu_c$ above $\nu_X$ for a long period of time (say, days),
one must have a small $\epsilon_B$ and/or a small $n$. Sometimes one
even needs to lower $\epsilon_e$ if one is reluctant to lower $n$. A
lower $\epsilon_B$ and a lower $\epsilon_e$ would lead to a larger
$E_K$ and hence, a lower $\eta_\gamma$. Since we do not know which
parameter is in operation, we fix $\epsilon_e = 0.1$ and $n=0.1$, and
search $\epsilon_B$ downwards to find the highest $\epsilon_B$ that
allows $\nu_m < \nu_X < \nu_c$ to be satisfied. In Table 4 we present
our search results for the regime II bursts.  The efficiency value
listed does not represent the real value. There is no obvious
reasons to argue why $\epsilon_e$ should be 0.1, why $n$ should be 0.1,
or why $\epsilon_B$ is not even lower.  Nonetheless, the value roughly
indicates how $E_K$ and $\eta_\gamma$ look when the regime II spectral
condition is satisfied. The results suggest that $\epsilon_B$ is
rather low, in most cases lower than $10^{-4}$. The estimated $E_K$ is
typically very high, and hence, the radiative efficiency is very
low. This suggests that either these GRBs are intrinsically
inefficient or the ambient densities of at least these GRBs are quite
low, say $n \ll 0.1~{\rm cm}^{-3}$.

%%%%%%%%%%%%%%%%%%%%%%%%%%%%%%%%%%%%%%%%%%%%%%%%%%%%%%%%%%%%%%%%%%%%%%

\section{Conclusions and discussion}
\label{sec:conclusion}

We present a detailed analysis on the prompt gamma-ray emission
and the early X-ray afterglow emission for a sample of 31 GRBs
detected by Swift and derived their radiative efficiencies. The sample
includes both long and short GRBs and both normal GRBs and softer
XRFs. This allows us to investigate how the GRB radiative efficiency
vary globally within different populations. We summarize our findings
in the following.

1. Due to the intrinsic limitation of the BAT instrument, it is very
difficult to derive spectral parameters of the prompt emission and to
extrapolate the BAT fluence to a broader band. We have developed a
method by making 
use the hardness ratio information to derive spectral parameters
assuming a Band function spectrum for all the bursts. This is probably
the best effort with the available BAT data. The involved
uncertainties is still large. It is also difficult to set errors
to the quantities. The errors of $\Gamma_1$ and $\Gamma_2$ of the Band
function are not accounted for in our analysis, but those of $E_0$
are included whenever possible. These errors are included to derive
the errors of $S_\gamma$, which also include the uncertainties in the
observations. Due to the intrinsic limitation of the method, the
errors of $S_\gamma$ are likely under-estimated.

2. We compare our prompt emission data with those of BATSE (Preece et
al. 2000) and HETE-2 (Lamb et al. 2005; Sakamoto et al. 2005) and
generally find consistent results. Swift extends the fluence to the
fainter regime and the hardness ratio to the softer regime with 
respect to BATSE. The derived properties of XRFs (including the 2:1
relative population between GRBs and XRFs) are generally
consistent with those derived from the HETE-2 data.

3. The shallow decay component commonly detected in X-ray afterglows
complicates the efficiency study. Previous analyses using late-time
X-ray afterglow data (say, 10 hr) inevitably over-estimated $E_K$ and
hence, under-estimated the GRB efficiency if the shallow decay is due
to energy injection. We define two characteristic time epoches,
the putative fireball deceleration time ($t_{\rm dec}$) and the epoch
when the shallow decay is over ($t_b$), to study the efficiency
problem. The efficiency derived at the former epoch
[i.e. $\eta_\gamma(t_{dec})$] is likely the true efficiency for the
models that interpret the shallow decay phase as due to energy injection
(Zhang et al. 2006; Nousek et al. 2006; Panaitescu et al. 2006), while
that derived at the
later epoch [i.e. $\eta_\gamma(t_b)$] is the true efficiency for some
models that invoke precursor injection (Ioka et al. 2006) or delayed
energy transition from the fireball to the circumburst medium
(Kobayashi \& Zhang 2006; Zhang \& Kobayashi 2005).

4. We investigate both the observationally defined gamma-to-X fluence
ratio ($R_{\gamma/X}$) and the theoretically defined efficiency
($\eta_\gamma$) at the two epochs. The former invokes direct
observables and therefore is subject to less uncertainties. The later
involve theoretical modeling, which depends on the uncertainties of
many unknown microphysics parameters. The error of the latter is
therefore difficult to quantify. Our approach is to fix model
parameters, and only derive the errors introduced from the
uncertainties of observations and data analyses. The errors of
calculated efficiencies are therefore under-estimated. Nonetheless,
some interesting features emerge from both the $R_{\gamma/X}$ analysis
and the $\eta_\gamma$ analysis. At $t_b$, a shallow correlation
between $R_{\gamma/X}$ or $\eta_\gamma$ and the hardness ratio is
found. This is consistent with previous findings (Soderberg et
al. 2004; Lloyd-Ronning \& Zhang 2004; Lamb et al. 2005) that XRFs
appear to be less efficient. However, the shallow correlation
disappears when the analysis is performed at $t_{dec}$. This suggests
that if the shallow decay is indeed due to energy injection, XRFs have
similar radiative efficiencies as normal GRBs. The apparent low
efficiency inferred using late time data must be attributed to some
other reasons.

The result has important implications for understanding the nature of
XRFs. In particular, it disfavors the model that interprets XRFs as
events similar to GRBs but having smaller Lorentz factor contrasts and
therefore lower radiative efficiencies (e.g. Barraud et al. 2005), if
the shallow decay phase is due to energy injection. It
suggests that XRFs are dim because the initial kinetic energy is also
low, and that they gain larger kinetic energies later through energy
injection.  While there is no straightforward reason for energy
injection in the radial direction (e.g. due to a long-lived central
engine or pile-up of slow ejecta) for XRFs, some geometric models of
XRFs indeed expect energy ``injection'' from the horizontal
directions. These models, such as the quasi-universal Gaussian-like
structured jet model (Zhang et al. 2004a; Dai \& Zhang 2005; Yamazaki
et al. 2004), invoke relativistic ejecta with variable luminosities
and possibly variable Lorentz factors in a wide range of
angles. While GRBs correspond to the cases when observers view the
bright core component (e.g. the line of sight is inside the typical
Gaussian angle), XRFs are those cases when observers view the off-axis
jet at a larger viewing angle. As long as the bulk Lorentz factor
of the ejecta is still relativistic in these directions, the prompt
emission of XRFs is weaker (because of the low energy in the
direction) but with a comparable efficiency as in the core direction,
since the initial kinetic energy in the direction is also low. Later
as the jet is decelerated, the observer in the off-axis direction
would progressively receive the energy contribution from the energetic
core so that the light curve shows an early shallow decay
(e.g. Kumar \& Granot 2003; Salmonson 2003). At later times, the
effective kinetic energy in the viewing direction is enhanced, leading
to an apparently high kinetic energy (and hence, an apparently low
radiative efficiency). Such a picture might be consistent with the
data. In fact, the long-term X-ray afterglow lightcurve of XRF 050416A
could be modeled by such a model (Mangano et al. 2006).

Other geometric models of XRFs have been also discussed in the
literature. The model involving sharp-edge jets viewed off beam
(Yamazaki et al. 2002) predicts an initial rising light curve and a
steep decay after reaching the peak due to sideways expansion, which
is not favored by the data. Granot et al. (2005) suggested a smoother
edge, which in effect is similar to the Gaussian jet model (Zhang et
al. 2004a). The two-component jet model (Zhang et al. 2004b; Huang et
al. 2004; Liang \& Dai 2004; Peng et al. 2005) could be consistent
with the data as long as the second component is not distinct enough to
result in noticeable light curve features that are not detected. Power
law structured jets (Zhang \& M\'esz\'aros 2002b; Rossi et al. 2002)
may also interpret XRFs (Jin \& Wei 2004; D'Alessio et al. 2006), but
the model predicts too large a XRF population (Lamb et al.  2005;
Zhang et al. 2004a). Finally, the varying opening angle model for XRFs
invoke very wide jets for XRFs (Lamb et al. 2005). However, there is
no straightforward reason to expect the significant change of $E_K$
(and hence, $\eta_\gamma$) at early times in this model.

It is possible that the shallow decay phase is not due to energy
injection. Numerical simulations (Kobayashi \& Zhang 2006) suggest
that the time scale for a fireball to transfer energy from the ejecta
to the medium is long. It could be that the observed shallow decay
phase reflects the epochs during which the energy transfer is going
on. A Poynting-flux-dominated flow may further extend the energy transfer
period further due to the inability of tapping kinetic energy of the
shell with the presence of a reverse shock (Zhang \& Kobayashi 2005).
If this is the case, $E_K(t_b)$ is the relevant kinetic energy to
define GRB efficiency, and XRFs are then indeed intrinsically
inefficient GRBs (e.g. Barraud et al. 2005). More detailed early
data and modeling are needed to reveal whether such a possibility
holds.

5. The absolute value of radiative efficiency is subject to the
uncertainties of afterglow parameters. By inspecting the spectral index
and the temporal decay index of the X-ray afterglows, we identify 22
bursts whose afterglow cooling frequency is below the X-ray band.  In this
regime the radiative efficiency is insensitive to parameters except
for $\epsilon_e$. Assuming $\epsilon_e = 0.1$, we find that
$\eta_{\gamma}(t_b)$ of most bursts are smaller than $10\%$, while
$\eta_{\gamma}(t_{\rm dec})$ ranges from a few percents to $>
90\%$. Some bursts have a low efficiency throughout, and they
correspond to those bursts whose X-ray afterglow light curve smoothly
join the prompt emission light curve without a distinct steep decay
component or an extended shallow decay component.

The standard internal shock models predict that the GRB efficiency is
only around $1\%$ (Kumar 1999; Panaitescu et al. 1999). Our results
indicate that some GRBs satisfy such a constraint. On the other hand, a
group of GRBs with an early shallow decay component challenge the
internal shock model if $E_{K}(t_{\rm dec})$ is the relevant quantity
to define the efficiency, as is required in most energy injection
models. Suggestions have been made in the literature to increase the
efficiency (e.g. Beloborodov 2000; Kobayashi \& Sari 2001; Fan et
al. 2004; Rees 
\& M\'esz\'aros 2005; Pe'er et al. 2005). For those models that
invoke $E_K(t_b)$ to define the efficiency (e.g. Ioka et al.  2006;
Kobayashi \& Zhang 2006), the data are consistent with the expectation
of the internal shock model. The typical efficiency in such a case is
$10\%$ or lower.

6.  The three short GRBs (050724, 050813 and 051221A) have similar
efficiencies as long GRBs at both $t_{\rm dec}$ and $t_b$, despite
their distinct progenitor systems (Gehrels et al. 2005; Fox et
al. 2005; Villasenor et al. 2005; Hjorth et al. 2005; Barthelmy et
al. 2005; Berger et al. 2005c; Bloom et al. 2006). Such a trend is
being verified by more and more short GRB data, and will be
reinforced by a larger sample of short GRBs in the future.

7. Although most of the X-ray afterglows in our sample are above the
cooling frequency, 9 GRBs do have a cooling frequency
higher than the X-ray band for a very long time, suggesting a very
small $\epsilon_B$ and/or a low medium density $n \ll 0.1~{\rm
cm}^{-3}$. An extremely low medium density may not be at odd for long
GRBs. If some of these long GRBs explode in superbubbles created by
precedented supernovae or GRBs, the ambient medium could reach such a
low density.

%%%%%%%%%%%%%%%%%%%%%%%%%%%%%%%%%%%%%%%%%%%%%%%%%%%%%%%%%%%%%%%%%%%%%%

\acknowledgments
We thank an anonymous referee for helpful comments. This work is
supported by NASA through grants NNG05GC22G, and NNG06GH62G (for BZ,
EL) and by the National Natural Science Foundation of China (for EL,
No. 10463001).

\clearpage

\begin{deluxetable}{llllllllllll}
%\rotate

\centering \tabletypesize{\scriptsize} \tablewidth{7in}

\tablecaption{\label{Gamma-ray}
Prompt Gamma-Ray data for the GRBs in our sample}

\tablecolumns{12}

\tablehead{

\colhead{GRB}      &

\colhead{$T_{90}$\tablenotemark{a}}    &

\colhead{$\Gamma_{PL}$\tablenotemark{b}}    &

\colhead{$\chi^2$\tablenotemark{b}}    &

\colhead{$HR_{\rm obs}$\tablenotemark{c}} &

\colhead{$\log S_{\gamma,obs}$ \tablenotemark{d}} &

\colhead{$\Gamma_1$\tablenotemark{e}} &

\colhead{$\Gamma_2$\tablenotemark{e}} &

\colhead{$E_p$\tablenotemark{e}} &

\colhead{$\chi^2$\tablenotemark{e}} &

\colhead{$\log S_\gamma$\tablenotemark{f}}&

\colhead{ $z^{ref.}$}
\\

\colhead{}      &

\colhead{(s)}    &

\colhead{}    &

\colhead{}    &

\colhead{} &

\colhead{(erg s$^{-1}$)} &

\colhead{} &

\colhead{} &

\colhead{(keV)} &

\colhead{} &

\colhead{(erg s$^{-1}$)}&

\colhead{}

}

\startdata
GRB050126 &    30.0&  -1.36$_{-0.15 }^{+0.15  }$&1.20 &    1.60(.23)&-6.07(.04)&-1.00&-2.30&  157$^{+ 129.}_{ -53.}$&1.19&   -5.63(.23)&1.290\tablenotemark{z1}\\
GRB050128 &    13.8&  NaN                      &     &    1.66(.14)&-5.29(.02)& -.71&-2.19&  118$^{+18.}_{-14.}$&0.87&   -4.87(.05)&\\
GRB050215B&    10.4&  -2.45$^{+0.73 }_{-1.02 }$&0.84 &     .84(.23)&-6.63(.07)&-1.00&-2.45&   $\sim $18&0.83&   $\sim $-6.24&\\
GRB050219A&    23.0&  NaN                      &     &    1.73(.11)&-5.38(.02)& -.32&-2.30&  102$^{+   8.}_{  -8.}$&0.91&   -5.01(.03)&\\
GRB050315 &    96.0&  -2.10$_{-0.09 }^{+0.09  }$&0.87 &     .93(.08)&-5.49(.02)&-1.28&-2.20&   37$^{+   8.}_{  -8.}$&0.88&   -5.10(.03)&1.950\tablenotemark{z2}\\
GRB050319 &    15.0&  -2.15$_{-0.21 }^{+0.20  }$&0.83 &     .99(.19)&-5.91(.05)&-1.25&-2.15&   $\sim$ 28&0.82&   $\sim $-5.51&3.240\tablenotemark{z3}\\
GRB050401\tablenotemark{g} &    38.0&  NaN &  &    1.50(.11)&-5.07(.02)&-1.15&-2.65&  132$^{+  16.}_{ -16.}$&0.96&   -4.74(.04)&2.900\tablenotemark{z4}\\
GRB050406 &     5.0&  -2.59$^{+0.38 }_{-0.45  }$&1.23 &     .74(.32)&-7.09(.10)&-1.00&-2.59&   $\sim$ 24                   &1.22&   $\sim $-6.70&\\
GRB050416A&     5.4&  -3.09$^{+0.22 }_{-0.24  }$&0.93 &     .44(.16)&-6.37(.06)&-1.00&-3.22&   $< 15$&0.90&   $\sim $-5.97& .654\tablenotemark{z5}\\
GRB050422 &    60.0&  -1.38$_{-0.22 }^{+0.22  }$&0.98 &    1.54(.28)&-6.21(.05)& -.90&-2.30&  123$^{+  99.}_{ -42.}$&0.79&   -5.83(.20)&\\
GRB050502B&    17.5&  -1.65$^{+0.14 }_{-0.15  }$&1.02 &    1.31(.18)&-6.33(.04)&-1.30&-2.30&  102$^{+  85.}_{ -36.}$&1.03&   -5.95(.10)&\\
GRB050525\tablenotemark{g} &    11.5&  NaN                      &     &    1.30(.03)&-4.81(.01)&-1.01&-2.72&   80$^{+   3.}_{  -3.}$&0.41&   -4.55(.01)& .606\tablenotemark{z6}\\
GRB050607 &    26.5&  -1.20$_{-0.07 }^{+0.07  }$&0.98 &    1.75(.16)&-6.22(.04)&-1.04&-2.00&  393$^{+ 368.}_{-132.}$&0.99&   -5.52(.27)& \\
GRB050712 &    48.0&  -1.48$_{-0.18 }^{+0.18  }$&0.99 &    1.42(.24)&-5.96(.05)&-1.00&-2.30&  126$^{+ 103.}_{ -44.}$&1.02&   -5.57(.18)&\\
GRB050713A\tablenotemark{g}&    70.0&  NaN &  &    1.39(.09)&-5.28(.02)&-1.12&-2.30&  312$^{+  50.}_{ -50.}$&1.22&   -4.72(.08)&\\
GRB050713B&    75.0&  -1.53$_{-0.17 }^{+0.17  }$&0.83 &    1.40(.22)&-5.34(.04)&-1.00&-2.30&  109$^{+  59.}_{ -32.}$&0.83&   -4.97(.11)&\\
GRB050714B&    87.0&  -2.59$^{+0.32 }_{-0.37  }$&0.88 &     .69(.43)&-6.24(.08)&-1.00&-2.59&   $\sim$20&0.86&  $\sim $ -5.85&\\
GRB050716 &    69.0&  NaN                      &     &    1.57(.10)&-5.20(.02)& -.87&-2.00&  119$^{+  19.}_{ -16.}$&0.88&   -4.70(.04)&\\
GRB050717\tablenotemark{g} &    70.0&  NaN &  &    1.63(.07)&-5.84(.03)&-1.12&-2.30& 1890$^{+1600.}_{-760.}$&0.78&   -4.85(.22)&\\
GRB050721 &    39.0&  -1.87$^{+0.16 }_{-0.17  }$&1.04 &    1.06(.25)&-5.51(.06)&-1.12&-2.05&   45$^{+  23.}_{ -35.}$&1.04&   -5.08(.09)&\\
GRB050724\tablenotemark{h} &     3.0&  -2.11$^{+0.24 }_{-0.26  }$&0.84 &     .88(.22)&-5.93(.06)&-0.65&-2.11&   $\sim$25&0.83&   $\sim $-5.45& .258\tablenotemark{z7}\\
GRB050726 &    30.0&  -0.99$^{+0.21 }_{-0.20  }$&1.23 &    2.14(.43)&-5.70(.05)&-1.00&-2.30&  $>$984&1.20&   $>$-4.79&\\
GRB050801 &    20.0&  -1.95$^{+0.19 }_{-0.20  }$&1.16 &    1.03(.27)&-6.51(.07)&-1.40&-2.00&   $\sim 33$&1.14&   $\sim $-6.02&\\
GRB050802 &    20.0&  -1.54$_{-0.14 }^{+0.14  }$&0.93 &    1.37(.19)&-5.66(.04)&-1.12&-2.30&  118$^{+  77.}_{ -40.}$&0.95&   -5.27(.13)&1.710\tablenotemark{z8}\\
GRB050803 &   150.0&  -1.43$_{-0.11 }^{+0.11  }$&0.97 &    1.51(.15)&-5.65(.03)&-1.05&-2.30&  150$^{+  68.}_{ -38.}$&0.97&   -5.24(.12)&\\
GRB050813 &      .6&  -1.42$_{-0.39 }^{+0.39  }$&1.35 &    1.66(.65)&-7.37(.11)& -.40&-2.30&   86$^{+ 101.}_{ -59.}$&1.29&   -7.03(.27)&\\
GRB050814 &    65.0&  -1.85$_{-0.18 }^{+0.18  }$&1.09 &    1.10(.20)&-5.74(.05)&-1.23&-2.77&   58$^{+  28.}_{ -16.}$&1.05&   -5.48(.05)&\\
GRB050819 &    40.0&  -2.64$^{+0.29 }_{-0.32  }$&0.99 &     .61(.22)&-6.45(.07)&-1.00&-2.64&   $<$15                  &1.04&   $<$-6.06(.07)&\\
GRB050820A&   270.0&  -1.24$_{-0.17 }^{+0.17  }$&1.26 &    1.73(.16)&-5.08(.02)&-1.00&-2.25&  284$^{+  82.}_{ -55.}$&1.27&   -4.52(.12)&2.612\tablenotemark{z9}\\
GRB050822 &   102.0&  -2.48$_{-0.15 }^{+0.15  }$&0.97 &     .73(.10)&-5.58(.03)&-1.00&-2.48&   $\sim 36$                  &0.96&  $\sim $ -5.19&\\
GRB050826 &    45.0&  -1.12$^{+0.29 }_{-0.28  }$&0.81 &    1.76(.43)&-6.35(.07)& -.93&-2.30&  $>240$&0.81&   $>$-5.82&\\
GRB051221A\tablenotemark{g}&     1.4&  NaN &  &    1.53(.08)&-5.94(.01)&-1.08&-2.30&  402$^{+  93.}_{ -72.}$&1.06&   -5.27(.11)& .547\tablenotemark{z10}\\
\enddata

\tablenotetext{a}{\ GRB duration in 15-150 keV.}

\tablenotetext{b}{\ The power law index and the reduced $\chi^2$ of
the best fit to the BAT data.}

\tablenotetext{c}{\ The Hardness ratio is calculated by gamma-ray
fluence in 50-100 keV band to that in 25-50 keV band.}

\tablenotetext{d}{\ The logarithm of the observed Gamma-ray fluence
and its error in the 15-150 keV band.}

\tablenotetext{e}{\ The spectral parameters derived from the best
Band-function fit with the constraint of $HR^{obs}=HR^{mod}$, except
for those bursts with marks. The errors of $E_p$ (in 90\% confidence
level) are derived from the best fits with Xspec package by fixing
both $\Gamma_1$ and $\Gamma_2$ .}

\tablenotetext{f}{\ Logarithm of extrapolated gamma-ray fluence in
$1-10^4$ keV band.}
% with the spectral parameters of the Band-function. }

%\tablenotetext{g}{The spectral parameters are taken from {\em HETE-2}
%data(Nakagawa et al. 2005)}

\tablenotetext{g}{The spectral parameters are taken from the {\em
Konus}-Wind data (Golenetskii et al 2005a-d; Krimm et al. 2006)}

\tablenotetext{h}{\ Most of the prompt emission was in one peak with
0.25s  and therefore the burst was considered as a short
burst(Barthelmy et al. 2005).}

Redshift references: (z1)Berger et al. (2005a), (z2) Berger et
al. (2005b), (z3)Fynbo et al. (2005a), (z4)Fynbo et al. (2005b),
(z5)Cenko et al. (2005), (z6)Foley et al. (2005), (z7)Berger et
al. (2005c), (z8)Fynbo et al. (2005c), (z9)Ledoux et al. (2005), (z10)
Berger \& Soderberg (2005).
\end{deluxetable}

\clearpage
\begin{deluxetable}{lllllllll}
%\rotate
\tabletypesize{\scriptsize} \tablewidth{7in}

\tablecaption{\label{X-Ray} Observations and Fitting Results of X-ray
Afterglows}

\tablecolumns{9}

\tablehead{

\colhead{GRB}      &

\colhead{$\beta$\tablenotemark{a}}    &

\colhead{$\alpha_1$\tablenotemark{b}} &

\colhead{$\alpha_2$\tablenotemark{b}}       &

\colhead{$\log(t_b/s)$\tablenotemark{b}}   &

\colhead{$\log S_{x,t_{\rm dec}}$\tablenotemark{c}}     &

\colhead{$\log S_{x,{t_b}}$\tablenotemark{c}} &

\colhead{$\log S_{x,1h}$\tablenotemark{c}}  & \colhead{$\log S_{x,10h}$\tablenotemark{c}} \\
}

\startdata
GRB050126 &    1.59( .38)&     .95( .27)&       --&   --&       -8.37(1.05)&       -8.23( .60)&       -8.27( .69)&       -8.22( .59)\\
GRB050128 &     .85( .12)&     .65( .10)&        1.25( .10)&    3.36( .10)&       -7.30( .16)&       -6.78( .18)&       -6.79( .19)&       -7.02( .23)\\
GRB050215B&     .51( .50)&     .82( .08)&       --&   --&       -8.31( .75)&       -7.79( .70)&       -7.98( .71)&       -7.80( .70)\\
GRB050219A&    1.02( .20)&     .59( .08)&       --&   --&       -8.40( .40)&       -7.26( .30)&       -7.64( .32)&       -7.23( .30)\\
GRB050315 &    1.50( .40)&     .01( .09)&         .74( .05)&    4.10( .14)&       -9.11( .08)&       -7.07( .10)&       -7.55( .08)&       -6.89( .13)\\
GRB050319 &    2.02( .47)&     .52( .03)&        1.77( .19)&    4.94( .13)&       -8.14( .12)&       -6.64( .18)&       -7.24( .12)&       -6.76( .12)\\
GRB050401 &     .98( .05)&     .68( .05)&        1.35( .05)&    3.74( .04)&       -7.01( .10)&       -6.42( .11)&       -6.44( .11)&       -6.65( .12)\\
GRB050406 &    1.37( .25)&     .51( .10)&       --&   --&      -10.11( .33)&       -8.05( .11)&       -9.20( .14)&       -8.71( .06)\\
GRB050416A&     .80( .29)&     .40( .05)&         .95( .05)&    3.18( .03)&       -8.51( .10)&       -7.68( .10)&       -7.61( .10)&       -7.55( .12)\\
GRB050422 &    2.23(1.09)&     .86( .04)&       --&   --&       -8.60( .20)&       -8.20( .16)&       -8.35( .17)&       -8.21( .16)\\
GRB050502B&     .81( .28)&     .86( .01)&       --&   --&       -8.38( .04)&       -7.97( .03)&       -8.12( .03)&       -7.98( .03)\\
GRB050525 &    1.07( .02)&    1.10( .05)&        1.45( .05)&    3.44( .05)&       -6.29( .11)&       -6.52( .12)&       -6.56( .12)&       -6.96( .13)\\
GRB050607 &     .77( .48)&    1.01( .05)&        2.46(1.65)&    5.11( .24)&       -7.97( .36)&       -8.06( .49)&       -7.99( .36)&       -8.00( .36)\\
GRB050712 &     .90( .06)&     .72( .01)&       --&   --&       -7.86( .17)&       -7.48( .14)&       -7.62( .15)&       -7.49( .14)\\
GRB050713A&    1.30( .07)&     .65( .10)&        1.15( .10)&    4.00( .04)&       -7.55( .27)&       -6.86( .27)&       -6.96( .27)&       -6.89( .28)\\
GRB050713B&     .70( .11)&     .30( .11)&         .97( .07)&    4.21( .23)&       -8.23( .20)&       -6.66( .24)&       -7.06( .20)&       -6.59( .29)\\
GRB050714B&    4.50( .70)&     .50( .06)&       --&   --&       -8.17( .32)&       -6.82( .26)&       -7.36( .27)&       -6.86( .26)\\
GRB050716 &     .33( .03)&    1.06( .07)&       --&   --&       -7.41( .38)&       -7.58( .31)&       -7.51( .32)&       -7.57( .31)\\
GRB050717 &    1.15( .10)&    1.49( .01)&       --&   --&       -6.63( .05)&       -8.00( .04)&       -7.47( .04)&       -7.96( .04)\\
GRB050721 &     .74( .15)&    1.28( .01)&       --&   --&       -6.78( .06)&       -7.60( .05)&       -7.30( .05)&       -7.58( .05)\\
GRB050724 &     .95( .07)&     .90( .0.1)&       --&   --&       -9.21( .33)&       -7.86( .07)&       -8.56( .14)&       -8.21( .06)\\
GRB050726 &     .94( .07)&     .91( .10)&        2.45( .10)&    4.06( .04)&       -7.45( .11)&       -7.30( .12)&       -7.28( .11)&       -7.96( .14)\\
GRB050801 &     .72( .54)&    1.10( .02)&       --&   --&       -7.70( .09)&       -7.99( .06)&       -7.89( .07)&       -7.99( .06)\\
GRB050802 &     .91( .19)&     .65( .10)&        1.65( .10)&    3.76( .04)&       -7.46( .10)&       -6.80( .11)&       -6.82( .10)&       -7.26( .14)\\
GRB050803 &     .71( .16)&     .61( .10)&       --&   --&       -7.58( .28)&       -6.77( .08)&       -7.04( .14)&       -6.65( .06)\\
GRB050813 &    2.42( .89)&     .70( .10)&        2.12( .10)&    2.26( .12)&       -9.40( .13)&       -9.29( .20)&      -10.69( .30)&      -11.81( .36)\\
GRB050814 &    1.08( .08)&     .62( .06)&       --&   --&       -8.63( .34)&       -7.51( .28)&       -7.97( .29)&       -7.59( .28)\\
GRB050819 &    1.18( .23)&    -.06( .06)&         .65( .10)&    4.26( .16)&      -10.85( .10)&       -8.20( .11)&       -8.89( .10)&       -8.04( .14)\\
GRB050820A&     .87( .09)&    1.18( .01)&       --&   --&       -6.07( .03)&       -6.46( .01)&       -6.27( .02)&       -6.45( .01)\\
GRB050822 &    1.60( .06)&     .46( .11)&         .95( .04)&    4.14( .31)&       -8.20( .29)&       -7.11( .33)&       -7.37( .29)&       -7.03( .38)\\
GRB050826 &    1.27( .47)&    1.10( .01)&       --&   --&       -7.41( .05)&       -7.70( .04)&       -7.60( .05)&       -7.70( .04)\\
GRB051221A&    1.04( .20)&    1.07( .03)&       --&   --&       -7.99( .11)&       -8.13( .06)&       -8.12( .06)&       -8.19( .05)\\
\enddata

\tablenotetext{a} {X-ray spectral index}

\tablenotetext{b} {$\alpha_1$ and $\alpha_2$ are the temporal decay
indices before and after the break time ($t_b$). If a light curve is
fitted by a simple power law only $\alpha_1$ is available}

\tablenotetext{c} {The values of $S_x$ at a given time are calculated
by the flux times the corresponding time, in units of erg cm$^{-2}$.}
\end{deluxetable}

\begin{deluxetable}{lllllllll}
%\rotate

\tabletypesize{\scriptsize}
\tablewidth{7in}
\label{Efficiency}
\tablecaption{The derived kinetic energies and gamma-ray Efficiencies
at $t_{dec}$ and $t_b$ for the GRBs in the spectral
regime $\nu_X>\max(\nu_m,\nu_c)$ by assuming $(\epsilon_e,
\epsilon_B)=(0.1,0.01)^{(1)}$
and $(\epsilon_e, \epsilon_B)=(0.1,0.0001)^{(2)} $}

\tablecolumns{10}

\tablehead{

\colhead{GRB}      &

\colhead{$\log E_{K}^{(1)}(t_{\rm dec})$}       &

\colhead{$\eta_\gamma^{(1)} (t_{\rm dec})$$^*$}   &

\colhead{$\log E_{K}^{(1)}(t_{b})$}     &

\colhead{$\eta_\gamma^{(1)}(t_b)$$^*$} &

\colhead{$\log E_{K}^{(2)}(t_{\rm dec})$}  &

\colhead{$\eta_\gamma^{(2)} (t_{\rm dec})$$^*$}&

\colhead{$\log E_{K}^{(2)}(t_{b})$}&

\colhead{$\eta_\gamma^{(2)}(t_b)$$^*$}
\\
\colhead{}      &

\colhead{(erg)}       &

\colhead{(100\%)}   &

\colhead{(erg)}     &

\colhead{(100\%)} &

\colhead{(erg)}  &

\colhead{(100\%)}&

\colhead{(erg)}&

\colhead{(100\%)}

} \startdata
GRB050126 &51.86(.84)&57.78(48.81)&53.60(.48)& 2.39( 2.84)&  52.26(.84)&35.27(45.68)&54.00(.48)&.97(1.16)\\
GRB050128 &54.48(.16)& 4.03(1.51)&55.01(.17)& 1.22(.50)&  54.49(.16)& 3.99(1.49)&55.02(.17)& 1.21(.50)\\
GRB050219A&52.84(.39)&57.21(22.20)&54.05(.30)& 7.57( 4.85)&  52.85(.39)&56.09(22.33)&54.07(.30)& 7.25(4.66)\\
GRB050315 &51.69(.06)&93.55(.96)&54.59(.08)& 1.79(.33)&  52.09(.06)&85.24( 2.00)&54.99(.08)&.72(.13)\\
GRB050319 &52.45(.10)&$\sim$ 70&55.59(.14)&$\sim0.2$&  52.85(.10)&$\sim$48&55.99(.14)&$\sim$.1\\
GRB050401 &55.05(.10)& 2.79(.70)&55.66(.11)&.70(.19)&  55.06(.10)& 2.76(.69)&55.66(.11)&.70(.18)\\
GRB050406 &50.67(.28)&$\sim$80&54.38(.09)&$\sim$0.1&  50.99(.28)&$\sim$66&54.69(.09)&$\sim$.04\\
GRB050416A&52.34(.10)&$\sim$5&53.18(.10)&$\sim$.8&  52.35(.10)& $\sim$5&53.18(.10)&$\sim $0.8\\
GRB050422 &51.80(.16)&68.99(12.56)&53.83(.13)& 2.01( 1.06)&  52.20(.16)&46.97(14.62)&54.23(.13)&.81(.43)\\
GRB050502B&53.40(.04)&3.99(.99)&53.84(.03)& 1.51(.37)&  53.41(.04)& 3.95(.98)&53.84(.03)& 1.50(.37)\\
GRB050525 &53.50(.11)&7.64(1.78)&53.45(.11)& 8.49(2.03)&  53.56(.11)& 6.61( 1.55)&53.52(.11)& 7.36( 1.78)\\
GRB050607 &53.82(.36)&4.11(4.06)&53.75(.49)& 4.78(5.86)&  53.82(.36)& 4.06( 4.02)&53.75(.49)& 4.72( 5.80)\\
GRB050712 &54.18(.14)&1.63(.86)&55.03(.14)&.24(.13)&  54.19(.14)& 1.61(.85)&55.03(.14)&.23(.12)\\
GRB050713A&52.92(.24)&68.02(12.41)&54.37(.24)& 7.07(3.76)&  53.19(.24)&53.84(14.18)&54.63(.24)&4.00( 2.20)\\
GRB050714B&50.80(.25)&$\sim$96&53.49(.21)& $\sim 4$&  51.20(.25)&$\sim 89$&53.89(.21)&$\sim 2$\\
GRB050717 &53.91(.05)&14.11( 6.38)&53.22(.04)&44.47(12.90)&  54.05(.05)&10.65(5.01)&53.36(.04)&36.74(12.14)\\
GRB050724 &50.40(.29)&69.23(14.68)&50.81(.07)&46.54(5.16)&  50.40(.29)&68.98(14.75)&50.81(.07)&46.26(5.15)\\
GRB050813 &51.03(.10)&44.81(16.69)&51.45(.16)&23.63(13.13)&  51.43(.10)&24.43(12.46)&51.85(.16)&10.96(7.10)\\
GRB050814 &52.17(.32)&67.78(16.47)&53.59(.27)& 7.38(4.31)&  52.25(.32)&63.80(17.42)&53.67(.27)& 6.26(3.70)\\
GRB050819 &49.95(.09)&$\sim 99$&53.02(.10)& $\sim$7&  50.11(.09)&$\sim$98.&53.18(.10)&$\sim 5$\\
GRB050822 &52.43(.23)&$\sim$69&54.59(.27)&$\sim$ 1.55(.94)&  52.83(.23)&$\sim $47&54.99(.27)&$\sim .6$\\
GRB050826 &53.01(.05)&$> 12$&53.80(.04)&$> 2$&  53.25(.05)& $> 7$&54.04(.04)&$> 1$\\
GRB051221A&51.92(.11)&32.56( 7.83)&51.90(.06)&33.57(6.54)&  51.96(.11)&30.61( 7.57)&51.94(.06)&31.59( 6.33)\\

\enddata

$^*$ The derived $\eta$ is insensitive to the redshift. For
those bursts whose redshifts are not available we take $z=2$ in
the calculation. The errors of $\eta$ are calculated by considering
only the errors from the extrapolated gamma-ray fluences and the
observed X-ray fluxes. Therefore, the errors of $\eta$ only reflect 
the observational errors. $E_K$ is sensitive to microphysics
parameters that are poorly constrained. The {\em true}
errors of $\eta$ should be significantly larger than what are reported
here.
\end{deluxetable}

\begin{deluxetable}{lllllll}
%\rotate
\tabletypesize{\scriptsize}
\tablewidth{4.5in}
\label{Efficiency}
\tablecaption{Derived kinetic energies and gamma-ray efficiencies at
$t_{dec}$ and $t_b$ for the GRBs in the spectral regime
$\nu_m<\nu_X<\nu_c$ with $\epsilon_e=0.1$}

\tablecolumns{7}

\tablehead{

\colhead{GRB} &

\colhead{$\epsilon_B$}  &

\colhead{$E_{K,52}(t_{\rm dec})$}&

\colhead{$\eta_\gamma (t_{\rm dec})$}&

\colhead{$E_{K,52}(t_{b})$}&

\colhead{$\eta_\gamma(t_b)$} &

\\
\colhead{} &

\colhead{}  &

\colhead{(erg)}&

\colhead{(100\%)}&

\colhead{(erg)}  &

\colhead{(100\%)} }

\startdata
GRB050215B&    $3\times 10^{-6}$&   53.65(    1.49)&    $\sim 1$&   54.17(    1.40)&  $\sim$0.4\\
GRB050713B&    $4\times 10^{-6}$&   52.01(     .35)&   90.68(    7.19)&   54.98(     .42)&    1.03(    1.02)\\
GRB050716 &    $3\times 10^{-7}$&   55.96(     .76)&     .20(     .36)&   55.09(     .62)&    1.48(    2.10)\\
GRB050721 &    $9\times 10^{-5}$&   54.55(     .10)&    2.11(     .66)&   53.54(     .08)&   18.29(    4.35)\\
GRB050726 &    $5\times 10^{-5}$&   53.22(     .16)&   $>$48&   54.38(     .18)&$>$6\\
GRB050801 &    $2\times 10^{-4}$&   52.84(     .15)&   $\sim 11$&   52.93(     .10)&   $\sim $9\\
GRB050802 &    $2\times 10^{-5}$&   52.95(     .16)&   29.60(    9.72)&   54.79(     .17)&     .61(     .29)\\
GRB050803 &    $6\times 10^{-6}$&   53.12(     .48)&   29.37(   23.78)&   54.79(     .14)&     .88(     .37)\\
GRB050820A&    $7\times 10^{-7}$&   56.50(     .04)&     .14(     .04)&   56.22(     .02)&     .27(     .07)\\

\enddata
\end{deluxetable}
\end{document}